\documentclass[cits]{PoS}

\usepackage{aas_macros}

\newcommand{\be}{\begin{equation}}
\newcommand{\ee}{\end{equation}}
\newcommand{\ba}{\begin{eqnarray}}
\newcommand{\ea}{\end{eqnarray}}

\newcommand\simgreater{\buildrel > \over \sim}
\newcommand\simless{\buildrel < \over \sim}

\newcommand{\Msol}{\mbox{$\mathrm{M}_{\odot}$}}
\newcommand{\rhonuc}{\mbox{$\rho_\mathrm{nuc}$}}
\newcommand{\eq}[1]{Eq.~(\ref{#1})}
\newcommand{\fig}[1]{Fig.~\ref{#1}}

\title{Superfluid Neutrons in the Core of the Neutron Star in Cassiopeia A}

\ShortTitle{Superfluidity in Cas A}

\author{
\speaker{Dany Page}\\
        Instituto de Astronom\'ia, 
        Universidad Nacional Aut\'onoma de M\'exico, \\
        Mexico, DF 04510, Mexico\\
        E-mail: \email{page@astro.unam.mx}}
\author{Madappa Prakash\\
        Department of Physics and Astronomy, 
        Ohio University,\\
        Athens, OH 45701-2979, USA\\
        E-mail: \email{prakash@harsha.phy.ohiou.edu}}
\author{James M. Lattimer\\
        Department of Physics and Astronomy, 
        State University of New York at Stony Brook,\\
        Stony Brook, NY 11794-3800, USA\\
       E-mail: \email{lattimer@mail.astro.sunysb.edu}}
\author{Andrew W. Steiner\\
        Institute for Nuclear Theory, University of Washington, \\
        Seattle, WA 98195, USA\\
        E-mail: \email{steiner3@uw.edu}}

\abstract{
The supernova remnant Cassiopeia A contains the youngest known neutron star
which is also the first one for which real time cooling has ever been observed. 
In order to explain the rapid cooling of this neutron star, we first
present the fundamental properties of neutron stars that control their
thermal evolution with emphasis on the neutrino emission processes and
neutron/proton superfluidity/superconductivity. 
Equipped with these results, we present
a scenario in which the observed cooling of the neutron star in
Cassiopeia A is interpreted as being due to the recent onset of
neutron superfluidity in the core of the star. The manner in which
the earlier occurrence of proton superconductivity determines the
observed rapidity of this neutron star's cooling is highlighted.
This is the first direct evidence that superfluidity and superconductivity occur at supranuclear densities within neutron stars.}

\FullConference{XXXIV edition of the Brazilian Workshop on Nuclear Physics,\\
		5-10 June 2011\\
		Foz de Igua\c{c}u, Parana state, Brasil}

\begin{document}

\section{Introduction: "neutron stars" do exist}

Neutron stars contain the densest form of cold matter observable in the Universe.
Larger energy densities are transiently reached in relativistic heavy ion collisions, but the resulting matter is 
extremely "hot". Black holes contain a 
 much denser form of matter, but their interior is not observable.
Two  simple arguments can convince us that such stars, very small and very dense, do exist.
First, consider the fastest known radio pulsar, "Ter5ad" (PSR J1748-2446ad) \cite{Hessels:2006fk}, and posit 
that the observed period
of its pulses, $P=1.39$ ms, is its rotational period. 
 Using {\em causality}, that is,   
 imposing that the
velocity at its equator is smaller than the speed of light $c$, one then obtains
\be
v_\mathrm{equator} = \Omega R = \frac{2\pi R}{P} < c
\;\;\; \Rightarrow \;\;\; 
R <  \frac{c P}{2 \pi} = 65 \,\mathrm{km} \; .
\ee
This value of 65 km for the radius $R$ is only a strict upper limit; detailed theoretical models 
indicate radii on the order
of 10 km.
Secondly,  
assuming that the star is {\em bound by gravity}, we can 
require that the gravitational
acceleration $a_\mathrm{gr} $ at the equator is larger than the centrifugal acceleration 
$a_\mathrm{cf}$ and obtain
\be
a_\mathrm{gr} = \frac{GM}{R^2} >
a_\mathrm{cf} = \Omega^2 R = \frac{4\pi^2 R}{P^2}
\;\;\; \mathrm{or} \;\;\;
\frac{M}{R^3} > \frac{4\pi^2}{GP^2}
\;\;\; \Rightarrow \;\;\; 
\overline{\rho} = \frac{M}{\frac{4}{3}\pi R^3}
> 8 \times 10^{13} \; \mathrm{g \, cm}^{-3} \; .
\ee
Obviously, Newtonian gravity is not accurate in 
this case, but we can nevertheless conclude that the
central density of these stars is comparable, or likely larger, than the nuclear density 
$\rhonuc \simeq 2.7 \times 10^{14}$ g cm$^{-3}$.
Theoretical models show that densities up to  $10 \rhonuc$ are possibly reachable \cite{Lattimer:2005fk}.
In short, a "neutron star" is a gigantic, and compressed, nucleus of the size of a city.

In what follows, we outline the basic properties of neutron stars
relevant for describing their thermal evolution with emphasis on the
neutrino emission processes and neutron/proton
superfluidity/superconductivity.  
This allows us to present simple
analytical models of the cooling of a neutron star in order to gain
physical insight. 
The results of these analytical models are
complemented by those of numerical simulations in which full
general relativity and the state of the art microphysics are employed.
Finally, an interpretation of the observed rapid cooling of the
neutron star in Cassiopeia A as due to the recent onset of neutron
superfluidity in it core is proffered. 
The role of proton superconductivity 
in determining the rapidity of Cas A's cooling is addressed. 

\section{Neutron Stars:  "pure neutron stars" do not exist}

A "pure neutron star", as 
 originally conceived by Baade \& Zwicky \cite{Baade-W.-and-Zwicky-F.:1934fk}
and Oppenheimer \& Volkoff \cite{Oppenheimer:1939uq} cannot really exist.
Neutrons in a ball should decay into protons through
\be
n \rightarrow p + e^- + \overline{\nu}_e \; .
\label{Eq:DU1}
\ee
This decay is possible for free neutrons as $m_n > m_p + m_e$. However, given the densities expected
within the neutron star interior, the relevant quantities are not the masses, but instead the chemical potentials 
$\mu_i$ ($i$ denoting the species) of the participants. 
The matter is degenerate as
typical Fermi energies are on the order of tens to hundreds of MeV's, whereas
the temperature drops below a few MeV within seconds after the birth of the neutron star in a core collapse
supernova \cite{Burrows:1986kx}.
Starting with a ball of almost degenerate neutrons, the decay of \eq{Eq:DU1} will generate a Fermi
sea of protons, electrons and anti-neutrinos. 
The interaction mean free paths of anti-neutrinos (and neutrinos) far exceed the size of the star
and these can be assumed to leave the star.
Thus, the decay will be possible until
\be
\mu_n = \mu_p + \mu_{e} \,,
\label{Eq:Chem}
\ee
where we have neglected $\mu_\nu$.
Under this condition, the inverse reaction
\be
p + e^-  \rightarrow n + \nu_e
\label{Eq:DU2}
\ee
also becomes energetically possible.
Hence, under equilibrium conditions, which a neutron star will reach within a few tens of seconds after its birth,
$\beta$ reactions such as \eq{Eq:DU1} and (\ref{Eq:DU2}) will adjust the chemical composition of matter
to that characteristic of  $\beta$ (or chemical) -equilibrium determined by \eq{Eq:Chem}.
A neutron star is not born as a "ball of neutrons" which may decay according to \eq{Eq:DU1}, but
from the collapse of the iron core of a massive star. It is thus born with an excess of protons so that 
the reaction \eq{Eq:DU2} initially dominates over \eq{Eq:DU1} in order to reduce the proton fraction 
and it is only when the neutrinos 
escape that the final cold $\beta$-equilibrium configuration, Eq.~(\ref{Eq:Chem}), is reached.

\begin{figure}
\begin{center}
\includegraphics[width=.65\textwidth]{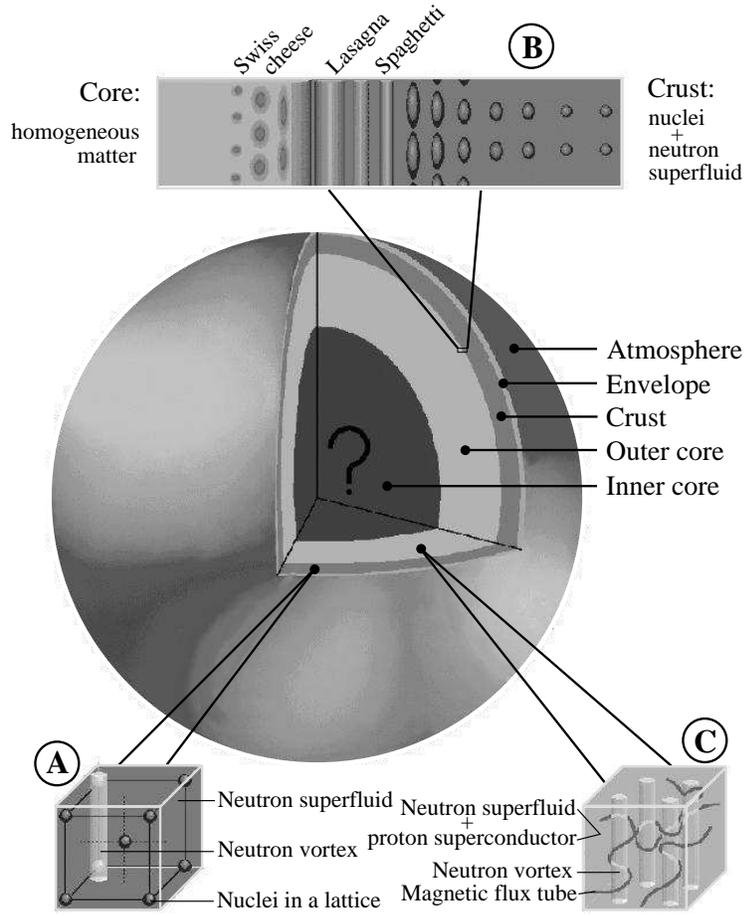}
\end{center}
\caption{Schematic illustration of the structure of a neutron star; figure taken  from \cite{Page:2006vn}.}
\label{Fig:NStar}
\end{figure}

Notice that once $\mu_e > m_\mu \simeq 105$ MeV, muons will appear, and be stably present with $\mu_\mu = \mu_e$.
The condition for the appearance of muons 
is fulfilled when the density is 
 slightly above $\rhonuc$.
Thus, in all processes we 
describe below, there will always be the possibility to replace electrons by muons when the 
density is large 
enough to allow for their presence.

Let us consider simple expressions for the chemical potentials~{\footnote{Relativistic expressions for $\mu_n$ and $\mu_p$
also exist, but are omitted here in the interest of simplicity.}}:
\ba
\mu_n = m_n + \frac{p_F(n)^2}{2 m_n} + V_n \qquad &{\rm and}& \qquad
\;\;\;\;
\mu_p = m_p + \frac{p_F(p)^2}{2 m_p} + V_p
\label{Eq:muN}
\\
\mu_e = \sqrt{m_e^2c^4 + p_F(e)^2 c^2} = p_F(e)c \qquad &{\rm and}& \qquad
\;\;\;\;
\mu_\mu = \sqrt{m_\mu^2c^4 + p_F(\mu)^2 c^2} \,,
\label{Eq:muL}
\ea
where $p_F(i)$ is the Fermi momentum of species $i$,
 and $V_n$ and $V_p$ are the mean-field energies of $n$ and $p$.
For the leptons, $V_e$ and $V_\mu$ are negligibly small, and we have considered that electrons, but not necessarily muons, are ultra-relativistic.
With a knowledge of $V_n$ and $V_p$, the two $\beta$-equilibrium relations \eq{Eq:Chem} and $\mu_\mu = \mu_e$ can be solved.
With four $\mu_i$'s and two equations, a unique solution is obtained by  imposing
charge neutrality
\be
n_p = n_e + n_\mu
\label{Eq:Qneutral}
\ee
and fixing the baryon density
\be
n_B = n_p + n_p \; .
\label{Eq:Baryon}
\ee
With the $n_i$'s and $\mu_i$'s known, one can calculate any thermodynamic potential, in particular the
pressure $P$ and energy density $\epsilon = \rho c^2$.
Varying the value of $n_B$ gives us the {\em equation of state} (EOS): $P =P(\rho)$.
Given an EOS, an integration of the Tolman-Oppenheimer-Volkoff (TOV) equations of hydrostatic
equilibrium provides us with a well defined model of a neutron star.

The potentials $V_n$ and $V_p$ in \eq{Eq:muN} turn out to be rapidly growing functions of density,
and one can anticipate that eventually reactions such as
\be
p + e^- \rightarrow \Lambda + \overline{\nu}_e
\;\;\;\;\;\;\;\; \mathrm{or/and} \;\;\;\;\;\;\;\;
n + e^- \rightarrow \Sigma^- + \overline{\nu}_e
\label{Eq:hyperons}
\ee
may produce hyperons.
Hyperons can appear, and be stable, once the corresponding $\beta$-equilibrium conditions
are satisfied, i.e., 
$\mu_n = \mu_\Lambda$ or/and $\mu_n + \mu_e = \mu_\Sigma^-$.
At the threshold, where $p_F(\Lambda) =0$ or $p_F(\Sigma^-) = 0$, one can expect that
$|V_{\Lambda}| \ll m_\Lambda$ and $|V_{\Sigma^-}| \ll m_{\Sigma^-}$
and thus
$\mu_\Lambda \simeq m_\Lambda$ and $\mu_{\Sigma^-} \simeq m_{\Sigma^-}$.
Since $m_\Lambda$ and $m_{\Sigma^-}$ are larger than the nucleon mass by only about 200 MeV
these hyperons\footnote{The $\Sigma^+$ is less favored as its $\beta$-equilibrium condition is
$ \mu_{\Sigma^+} = \mu_p = \mu_n - \mu_e$. Heavier baryon are even less favored, but cannot
{\em a priori} be excluded.}
are good candidates for an "exotic" form of matter in neutron stars.
Along similar lines, 
 the lightest mesons, pions and/or kaons, may also appear stably,
and can form meson condensates.
At even larger densities, the ground state of matter is likely to
be one of deconfined quarks.
All these possibilities
depend crucially on the strong interactions terms,  $V_n$ and $V_p$.
As we will not employ them in this chapter, we refer, e.g., to \cite{Page:2006vn} for more details
and entries to the original literature.
Figure~\ref{Fig:NStar} illustrates our present understanding (or misunderstanding) of the interior of a
neutron star, with a black question mark "?" in its densest part.
The outer part of the star, its {\em crust}, is described briefly in the following subsection.

When only nucleons, plus $e$'s and $\mu$'s as implied by charge neutrality and constrained by $\beta$-equilibrium,
 are considered, the EOS can be calculated with much more confidence than in the presence of "exotic" forms of matter. 
For illustrative puposes, we will employ the EOS of  Akmal, Pandharipande \& Ravenhall 
\cite{Akmal:1998qf} ("APR" hereafter) in presenting our results.

Although there is no evidence that any observed "neutron star" or pulsar 
might actually instead be a pure quark star, theory allows this possibility.  
Such a star would be nearly completely composed of a mixture of up, down 
and strange quarks, and would differ from a neutron star in that it would be 
self-bound rather than held together by gravity.

The reader can find a more detailed presentation and entries to the key literature in \cite{Page:2006vn}.

\subsection*{The Neutron Star Crust}
\label{Sec:Crust}

In the outer part of the star, where $\rho < \rhonuc$, a homogeneous liquid of nucleons is mechanically unstable (spinodal instablity).
Stability is, however, restored by the formation of nuclei, or nuclear clusters.
At the surface, defined as the layer where $P =0$, we expect the presence of an atmosphere, but there is 
the possibility of having a solid surface condensed by a sufficiently strong magnetic field \cite{Lai:2001uq}.
A few meters below the surface, ions are totally ionized by the increasing density (the radius of the first Bohr orbital will be larger than the inter-nuclear distance when $\rho \simgreater 10^4$ g cm$^{-3}$).
Matter then consists of a gas/liquid of nuclei immersed in a quantum liquid of electrons.
When $\rho \approx 10^6$ g cm$^{-3}$, $\mu_e$ is of the order of 1 MeV and the electrons become relativistic. 
From here on, the Coulomb correction to the electron gas EOS is negligible and electrons
form an almost perfect Fermi gas.
However, the Coulomb correction to the ion EOS is {\em not} negligible. From a gaseous state at the surface, ions will
progressively go through a liquid state and finally crystallize, at densities between $10^2$ up to $\sim 10^{10}$
g cm$^{-3}$ depending on the temperature (within the range of temperatures for which the neutron star is
thermally detectable).
With growing $\rho$, and the accompanying growth of $\mu_e$, it is energetically favorable to absorb electrons
into nuclei and, hence, the nuclear species expected to be present have a neutron fraction strongly growing
with $\rho$.
When $\rho \sim 4-8 \times 10^{11}$ g cm$^{-3}$ (the exact value depending on the assumed chemical 
composition), one reaches the {\em neutron drip} point at which 
the neutron density is so much larger than that of the proton that not all neutrons are bound to nuclei.
Matter then consists of a crystal of nuclei immersed in a perfect Fermi gas of electrons and a quantum liquid of 
dripped neutrons.
This region is usually called the {\em inner crust}.
In most of this inner crust, the dripped 
neutrons are predicted to form a superfluid (in a spin-singlet, 
zero orbital angular momentum, state $^1$S$_0$).
All neutron stars we observe are rotating; a superfluid cannot undergo rigid body rotation, but
it can simulate it by forming an array of vortices (in the core of which superfluidity is destroyed).
(See, e.g., \cite{Tilley:1990ys}.)
The resulting structure is illustrated in the inset A of Fig.~\ref{Fig:NStar}.

At not too high densities, nucleons are correlated at short distances by the strong interaction and anti-correlated
at larger distances by the Coulomb repulsion between the nuclei, 
the former producing spherical nuclei and the latter resulting in the crystallization of the matter. 
As $\rho$ approaches $\rhonuc$ the shape of the "nuclei" can undergo drastic changes: the nuclear attraction 
and Coulomb repulsion length-scales become comparable and the system is "frustrated". 
From spherical shapes, nuclei are expected to become elongated ("spaghettis"), form 2D structures
("lasagnas"), always surrounded by the 
neutron gas/superfluid which occupies an increasingly
small portion of the volume. 
Then the geometry is inverted, with the dripped 
neutrons confined into 2D, then 1D ("anti-spaghettis") and
finally 0D ("swiss cheese") bubbles. 
The homogeneous phase, i.e. the {\em core} of the star, is reached when 
$\rho = \rho_\mathrm{cc} \simeq 0.6 \rhonuc$
\footnote{$\rhonuc$ correspond to the density of symmetric nuclear matter, i.e. with a proton fraction
$x_p = 50$\%, and zero pressure, whereas in a neutron star crust at $\rho \sim \rhonuc$ one has $x_p \simeq 3-5$\%.}.
This "pasta" regime is illustrated in the inset B of Fig.~\ref{Fig:NStar} and is thought to resemble  a liquid crystal \cite{Pethick:1998fk}.

\section{Neutrino Emission Processes}
\label{Sec:Neutrinos}

The thermal evolution of neutron stars
with ages $\simless 10^5$ yrs is driven by neutrino emission.
Here we describe the dominant processes.
The simplest neutrino emitting processes, \eq{Eq:DU1} and \eq{Eq:DU2}, are  
\be
\left\{
\begin{array}{lcl}
n & \longrightarrow & p + e^- + \overline{\nu}_e
\\
p + e^-  & \longrightarrow & n + \nu_e \,,
\end{array}
\right.
\label{Eq:DUN}
\ee
and are generally referred to as the  {\em direct Urca} ("DU") cycle. 

By the condition of $\beta$-equilibrium the cycle naturally satisfies energy conservation, but momentum
conservation is much more delicate.
Due to the high degeneracy, all participating particles have momenta $p(i)$ equal (within a small $T \ll T_F$
correction) to their Fermi momenta $p_F(i)$.
As $p_F(i) \propto n_i^{1/3}$ and $n_p \sim n_e \ll n_n$, momentum conservation is not {\em a priori}
guaranteed.
It is easy to see that, in the absence of muons and hence with $n_p = n_e$, the "triangle rule" for
momentum conservation requires that $x_p > 1/9 \simeq 11$\%, whereas at $\rho \sim \rhonuc$ we have
$x_p \simeq 5$\%.
In the presence of muons, which appear just above $\rhonuc$, the condition is stronger and one needs
$x_p$ larger than about 15\% \cite{Lattimer:1991kx}.
The proton fraction $x_p$ grows with density, its growth being directly determined by the growth of the nuclear symmetry energy,
so that the critical proton fraction for the DU process is likely reached at some supra-nuclear density \cite{Lattimer:1991kx}.
For the EOS of APR, the corresponding critical neutron star mass for the allowance of the DU
process is $1.97 \Msol$, but other EOS's predict smaller values.

\begin{table}
\begin{tabular}{llcc} 
\hline 
     Name           &               Process              &       Emissivity       &  \\ 
                    &                                        &   (erg cm$^{-3}$ s$^{-1}$) &               \\
\hline 
\parbox[c]{3.4cm}{Modified Urca\\ (neutron branch)} &
\rule[-0.4cm]{0.02cm}{0.85cm}
$\begin{array}{l} n+n \rightarrow n+p+e^-+\bar\nu_e \\ n+p+e^- \rightarrow n+n+\nu_e \end{array}$  & 
$\sim 2\!\!\times\!\! 10^{21} \: R \: T_9^8$ & Slow \\
\parbox[c]{3.4cm}{Modified Urca\\ (proton branch)}  &
\rule[-0.4cm]{0.02cm}{0.85cm}
$\begin{array}{l} p+n \rightarrow p+p+e^-+\bar\nu_e \\ p+p+e^- \rightarrow p+n+\nu_e \end{array}$  & 
$\sim 10^{21} \: R \: T_9^8$ & Slow \\
Bremsstrahlungs          &
$\begin{array}{l} n+n \rightarrow n+n+\nu+\bar\nu \\ n+p \rightarrow n+p+\nu+\bar\nu \vspace{-0.0cm} \\
    p+p \rightarrow p+p+\nu+\bar\nu \end{array}$                                     & 
$\sim 10^{19} \: R \: T_9^8$  & Slow \\
\parbox[c]{3.5cm}{Cooper pair}          &
$\begin{array}{l}    n+n \rightarrow [nn] +\nu+\bar\nu \\ p+p \rightarrow [pp] +\nu+\bar\nu \end{array}$  & 
$\begin{array}{l} \sim 5\!\!\times\!\! 10^{21} \: R \: T_9^7 \\ \sim 5\!\!\times\!\! 10^{19} \: R \: T_9^7 \end{array}$ & Medium \\
\parbox[c]{3.4cm}{Direct Urca\\ (nucleons)} &
\rule[-0.4cm]{0.02cm}{0.85cm}
$\begin{array}{l} n \rightarrow p+e^-+\bar\nu_e \\ p+e^- \rightarrow n+\nu_e \end{array}$              & 
$\sim 10^{27} \: R \: T_9^6$ & Fast \\
\parbox[c]{3.4cm}{Direct Urca\\ ($\Lambda$ hyperons)} &
\rule[-0.4cm]{0.02cm}{0.85cm}
$\begin{array}{l} \Lambda \rightarrow p+e^-+\bar\nu_e \\ p+e^- \rightarrow \Lambda+\nu_e \end{array}$              & 
$\sim 10^{27} \: R \: T_9^6$ & Fast \\
\parbox[c]{3.4cm}{Direct Urca\\ ($\Sigma^-$ hyperons)} &
\rule[-0.4cm]{0.02cm}{0.85cm}
$\begin{array}{l} \Sigma^- \rightarrow n+e^-+\bar\nu_e \\ n+e^- \rightarrow \Sigma^-+\nu_e \end{array}$              & 
$\sim 10^{27} \: R \: T_9^6$ & Fast \\%
$\pi^-$ condensate &$n+<\pi^-> \rightarrow n+e^-+\bar\nu_e$  &
$\sim 10^{26} \: R \: T_9^6$  & Fast \\
$K^-$ condensate   &$n+<K^-> \rightarrow n+e^-+\bar\nu_e$  &
$\sim 10^{25} \: R \: T_9^6$  & Fast \\
\hline 
\parbox[c]{3.4cm}{Direct Urca cycle\\ (u-d quarks)} &
\rule[-0.4cm]{0.02cm}{0.85cm}
$\begin{array}{l} d \rightarrow u+e^-+\bar\nu_e \\ u+e^- \rightarrow d+\nu_e \end{array}$              & 
$\sim 10^{27} \: R \: T_9^6$ & Fast \\
\parbox[c]{3.4cm}{Direct Urca cycle\\ (u-s quarks)} &
\rule[-0.4cm]{0.02cm}{0.85cm}
$\begin{array}{l} s \rightarrow u+e^-+\bar\nu_e \\ u+e^- \rightarrow s+\nu_e \end{array}$              & 
$\sim 10^{27} \: R \: T_9^6$ & Fast \\
\hline 
\end{tabular}
\caption{
A sample of neutrino emission processes. 
$T_9$ is temperature $T$ in units of $10^9$ K and the $R$'s are control factors
to include the suppressing effects of pairing (see \S\protect\ref{Sec:Pairing}).
}
\label{Tab:Nu}
\end{table}

At densities below the threshold density for the DU process, a variant of this process, 
the {\em modified Urca} (MU) process 
\be
\left\{
\begin{array}{lcl}
n + n' & \longrightarrow & p + n' + e^- + \overline{\nu}_e
\\
p + n' + e^-  & \longrightarrow & n + n' + \nu_e
\end{array}
\right.
\label{Eq:MU}
\ee
can operate as advantage is taken of neighboring particles in the medium \cite{Friman:1979bh}.
A bystander neutron $n'$ can take or give the needed momentum for momentum conservation.
The processes in \eq{Eq:MU} show the {\em neutron branch} of the MU process and in the {\em proton branch} 
$n'$ is replaced by a proton $p'$. 
As it involves the participation of five degenerate particles, the MU process is much less efficient than the DU process.
Unlike the DU processes which require sufficient amount of protons, 
both branches of the MU process operate at any density when neutrons and protons are present.

In the presence of hyperons, DU and MU processes which are obvious generalizations of the nucleon-only
process can also occur \cite{Prakash:1992vn}. 
When they appear, the $\Lambda$'s have a density much smaller than that of the neutron
and hence
a smaller Fermi momentum. Consequently, momentum conservation in the DU cycle
$p + e^- \rightarrow \Lambda + \overline{\nu}_e$ and 
$\Lambda \rightarrow  p + e^- + \nu_e$ is easily satisfied, requiring an $x_\Lambda$ of only $\sim 3$\%.
Notice that if the nucleon DU process is kinematically forbidden, the hyperon DU process
$\Sigma^- \rightarrow n + e^- + \overline{\nu}_e$ with
$n + e^- \rightarrow \Sigma^- + \nu_e$ is also kinematically forbidden, whereas the no-nucleon DU process
$\Sigma^- \rightarrow \Lambda + e^- + \overline{\nu}_e$ together with
$\Lambda + e^- \rightarrow \Sigma^- + \nu_e$ may be possible and require very low threshold fractions
$x_\Lambda$ and $x_\Sigma^-$.

In deconfined quark matter, DU processes such as 
$u + e \rightarrow d + \overline{\nu}_e$ and  
$d \rightarrow u + e + \nu_e$ are possible \cite{Iwamoto:1982ys}.
Reactions in which the $d$ quark is replaced by an $s$ quark can also occur  in the case $s$ quarks are present.

In the presence of a meson condensate copious neutrino emission in processes such as
\be
\left\{
\begin{array}{lcl}
n + e^- & \longrightarrow & n + \pi^- + \nu_e
\\
n + \pi^- & \longrightarrow & n + e^- + \overline{\nu}_e
\end{array}
\right.
\label{Eq:pionDU}
\ee
and 
\be
\left\{
\begin{array}{lcl}
n + e^- & \longrightarrow & n + K^- + \nu_e
\\
n + K^- & \longrightarrow & n + e^- + \overline{\nu}_e \, 
\end{array}
\right. 
\label{Eq:kaonDU}
\ee
occur \cite{Maxwell:1977zr,Brown:1988ly}.
As the meson condensate is a macroscopic object there is no restriction arising from momentum
conservation in these processes.

Finally, another class of processes, bremsstrahlung, is possible through neutral currents \cite{Flowers:1975qf}.   
Reactions such as
\ba
n + n' \longrightarrow n + n' + \nu_e + \overline{\nu}_e
\\
p + p' \longrightarrow p + p' + \nu_e + \overline{\nu}_e
\\
n + p' \longrightarrow n + p' + \nu_e + \overline{\nu}_e
\label{Eq:NBrem}
\ea
are less efficient, by about 2 orders of magnitude, than the MU processes, but may
make some contribution in the case that the MU process is suppressed by pairing of neutrons or
protons as we will see in \S\ref{Sec:Pairing}.

In Table~\ref{Tab:Nu} we list these processes with order of magnitude estimates of their neutrino emissivities.
Most noticeable is the clear distinction between processes involving 5 degenerate fermions with a
$T^8$ dependence, which are labeled as "slow", and those with only 3 degenerate fermions
with a $T^6$ dependence, which are several orders of magnitude more efficient and labeled as "fast".
The difference in the $T$ dependence is important and simply related to phase space arguments
which are outlined below.
The Cooper pair process \cite{Flowers:1976vn,Voskresensky:1987uq} will be described in \S\ref{Sec:Pairing}.

The reader can find a detailed description of neutrino emission processes in \cite{Yakovlev:2001dq}
and an alternative point of view in \cite{Voskresensky:2001cr}.

\subsection*{Temperature dependence of neutrino emission}
\label{Sec:nuT}

We turn now to briefly describe how the specific temperature dependence of the neutrino processes described
above emerges.
Consider first the simple case of the neutron $beta$ decay.
The weak interaction is described by the Hamiltonian ${\cal H}_I = (G_F/\sqrt{2}) B_\mu L^\mu$, where
$G_F$ is Fermi's constant, and 
$L^\mu = \overline{\psi}_e \gamma^\mu (1-\gamma_5) \psi_\nu$
and
$B_\mu = \overline{\psi}_p \gamma_\mu (C_V 1-C_A \gamma_5) \psi_n$ 
are the lepton and baryon currents, respectively.
In the non relativistic approximation, one has
$B^0 = \cos \theta_c \Psi^\dagger_p  \Psi_n$ and
$B^i = -\cos \theta_c  \, g_A \; \Psi^\dagger_p \, \sigma^i \, \Psi_n$
where $\theta_c$ is the Cabibbo angle and $g_A$ the axial-vector coupling.
Fermi's Golden rule gives us for the neutron decay rate
\be
W_{i \rightarrow f} =
\int \!\!\!\! \int \!\!\!\! \int
\frac{d^3 p_\nu}{(2 \pi)^3} \frac{d^3 p_e}{(2 \pi)^3}
\frac{d^3 p_p   }{(2 \pi)^3}
 (2 \pi)^4 \delta^4(P_f-P_i) \cdot |M_{fi}|^2 \, ,
\ee
i.e., a sum of $(2 \pi)^4 \delta^4(P_f-P_i) \cdot |M_{fi}|^2$ over the phase space of all final states 
$f=(\vec{p}_{\overline{\nu}},\vec{p}_e,\vec{p}_p)$.
The integration gives the well known result:
$W_\beta = \frac{1}{2 \pi^3}[G_F^2 \cos^2 \theta_c (1+3 g_A^2) m_e^5 c^4] w_\beta$,
where $w_\beta \sim 1$ takes into account small Coulomb corrections.
This gives us the neutron mean life, $\tau_n \simeq 15$ min., or, measuring $\tau_n$, a measurement of
$G_F$ (modulo $\cos \theta_c$ and $w_\beta$).

\begin{figure*}[t]
\begin{center}
\includegraphics[width=.9\textwidth]{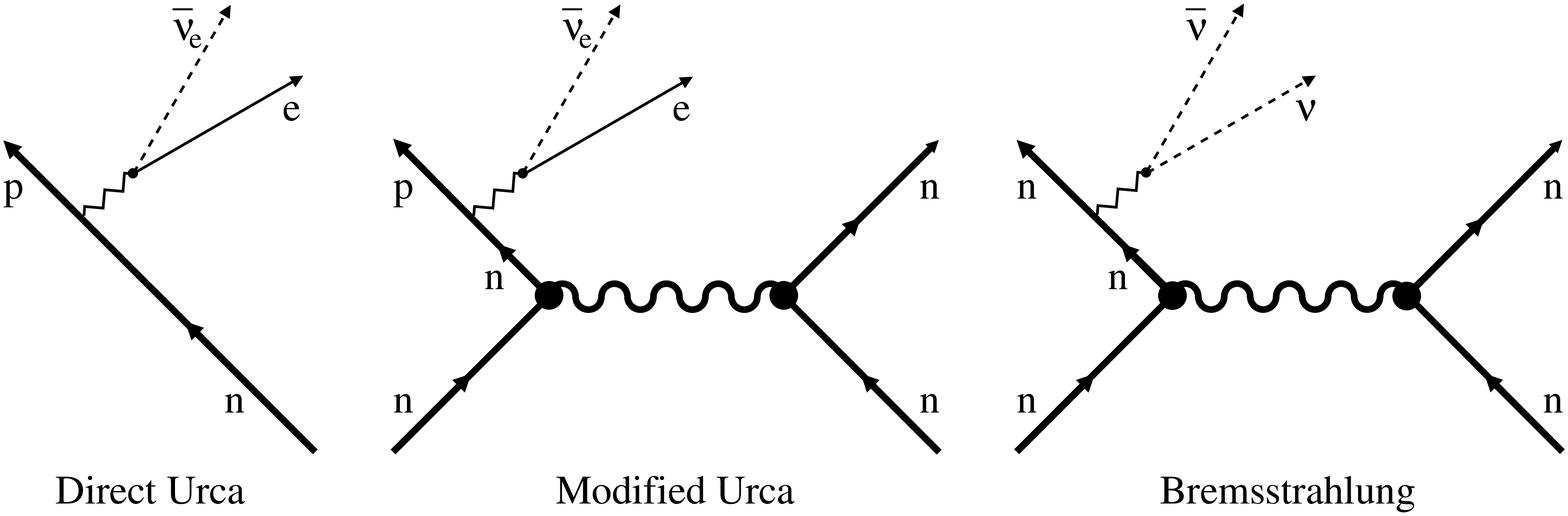}
\end{center}
\caption{Feynman diagrams for the indicated neutrino emitting processes.}
\label{Fig:Feynman}
\end{figure*}

The emissivity $\epsilon^\mathrm{DU}$ of the DU process 
(the Feynman diagram for this process is shown in \fig{Fig:Feynman})
can be obtained by the same method as above leading to
\be
\epsilon^\mathrm{DU} =
\int \!\!\!\! \int \!\!\!\! \int \!\!\!\! \int 
\frac{d^3 p_{\overline{\nu}}}{(2 \pi)^3} \frac{d^3 p_e}{(2 \pi)^3}
\frac{d^3 p_p   }{(2 \pi)^3} \frac{d^3 p_n}{(2 \pi)^3}
(1-f_e) (1-f_p) f_n \cdot  (2 \pi)^4 \delta^4(P_f-P_i) |M_{fi}|^2 \cdot E_\nu
\label{Eq:E_DU}
\ee
with an extra factor $E_\nu$ for the neutrino energy and the phase space sum now includes the initial $n$.
The $f_i$ terms, $f_i$ being the Fermi-Dirac distribution for particle $i$ at temperature $T$,
take into account:
(1) the probability to have a $n$ in the initial state, $f_n$, and
(2) the probabilities to have available states for the final $e$ and $p$, denoted by $(1-f_e)$ and $(1-f_p)$, respectively.
We do not introduce a Pauli blocking factor $(1-f_{\overline{\nu}})$ for the anti-neutrino as it is assumed
to be able to freely leave the star (i.e., $f_{\overline{\nu}}=0$).
When performing the phase space integrals, each degenerate fermion gives us a factor $T$, as particles are 
restricted to be within a shell of thickness $k_BT$ of their respective Fermi surfaces.  The  anti-neutrino phase space
gives a factor $T^3$. The factors $E_\nu \sim T$ and the delta function $\delta^4(P_f-P_i)$ gives a factor
$T^{-1}$ (it cancels one of the $T$'s from the degenerate fermions).
Altogether, we find that
\be
\epsilon^\mathrm{DU} \propto
T^3 \cdot T  \cdot T  \cdot T \cdot \frac{1}{T} \cdot (1)^2 \cdot T = T^6 \,,
\label{Eq:DU_T}
\ee
where the $(1)^2$ factor emphasizes that the squared matrix element $|M_{fi}|^2$ is T-independent. 
An explicit expression for the neutrino emissivity for the DU process can be found in \cite{Lattimer:1991kx} . 

Figure  \fig{Fig:Feynman} shows 
a Feynman diagram for the MU process.  There are two more such diagrams
in which the weak interaction vertex is attached to one of the two incoming legs.
In this case, the $T$-power counting gives
\be
\epsilon^\mathrm{MU} \propto
T^3 \cdot T  \cdot T  \cdot T \cdot T \cdot T \cdot \frac{1}{T} \cdot (1)^2 \cdot T = T^8 \,,
\label{Eq:MU_T}
\ee
where the $|M_{fi}|^2$ now involves a strong interaction vertex, the wavy line in \fig{Fig:Feynman},
but is still $T$-independent. 
Notice that in the MU case, the internal neutron is off-shell by an amount $\simeq \mu_e$ which does not
introduce any extra $T$-dependence as we are working in the case 
$E_F(e) \simgreater 100$ MeV $\gg T$.
Reference \cite{Friman:1979bh} contains the expression from which neutrino emissivity from the MU process can be calculated.
Considering finally the $n-n$ bremsstrahlung process. 
One diagram is shown in \fig{Fig:Feynman} and there are three more diagrams with the weak interaction vertices attached 
to the other three external lines. The $T$-power counting now gives
\be
\epsilon^\mathrm{Br} \propto
T^3 \cdot T^3  \cdot T  \cdot T \cdot T \cdot T \cdot \frac{1}{T} \cdot \left(\frac{1}{T}\right)^2 \cdot T = T^8
\ee
with two $T^3$ factors for the neutrino pair and a $(T^{-1})^2$ from the matrix element:
the intermediate neutron is almost on-shell, with an energy deficit $\sim T$, and its propagator gives us
a $T^{-1}$ dependence for $M_{fi}$. 
A working expression for the bremsstrahlung process can be found in \cite{Friman:1979bh}. 
 
\section{Neutron Star Cooling}

The basic features of the cooling of a neutron star are best illustrated
by considering the energy balance equation for the star in its
Newtonian formulation{\footnote{Numerical results to be shown later include full general relativistic effects.}}:
\be
\frac{dE_\mathrm{th}}{dt} = C_\mathrm{v} \frac{dT}{dt}
                   = -L_\nu - L_\gamma + H \, ,
\label{Eq:Cooling}
\ee
where $E_\mathrm{th}$ is the thermal energy content of the star, $T$ its
internal temperature, and $C_\mathrm{v}$ its total specific heat. The two
energy sinks are neutrino luminosity $L_\nu$, described in
\S\ref{Sec:Neutrinos}, and the surface photon luminosity $L_\gamma$,
discussed in \S\ref{Sec:Envelope}.  The source term $H$ would include
heating mechanisms as, e.g., magnetic field decay, which we will not consider here.

\subsection{Specific Heat}
\label{Sec:Cv}

\begin{figure*}[b]
\begin{center}
\includegraphics[width=.4\textwidth]{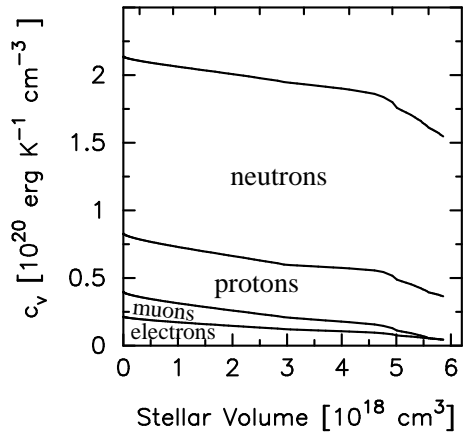}
\end{center}
\caption{Cumulative specific heats of e, $\mu$, p, and n vs. stellar volume in
the core of a 1.4 $M_\odot$ star built using the EOS of APR, at temperature $T=10^9$ K.
Nucleons are assumed to be unpaired.
No hyperons or quarks are present in the EOS.
This figure is adapted from \cite{Page:2004zr}.}
\label{Fig:Cv}
\end{figure*}

The dominant contributions to the specific heat $C_\mathrm{v}$ come from the core, constituting 
more than 90\% of the total volume, whose constituents are quantum liquids of leptons,
baryons, mesons, and possibly deconfined quarks at the highest densities.
Hence one has
\be
C_\mathrm{V} = \sum_i C_{\mathrm{V}, \, i}
\;\;\;\;\;\;
\mathrm{with}
\;\;\;\;\;\;
 C_{\mathrm{V}, \, i} = \int \!\!\!\! \int \!\!\!\! \int c_{\mathrm{v},\, i} \, d\mathrm{v}
\;\;\;\;\;\;
(i = \mathrm{e, \, \mu, \, n, \, p, \, hyperons, \, quarks, ...}) \,,
\label{Eq:CvTot}
\ee
where $c_{\mathrm{v},\, i}$ is the specific heat, per unit volume, of component $i$.
For normal (i.e., unpaired) degenerate fermions, the standard Fermi liquid
result \cite{Baym:2004nx}
\be
c_{\mathrm{v} \; i} = N(0) \frac{\pi^2}{3} k_B^2 T
\;\; \mathrm{with} \;\;
N(0) = \frac{m_i^* p_{F \, i}}{\pi^2 \hbar^3}
\label{Eq:Cv}
\ee
can be used, where $m^*$ is the fermion's effective mass.
In Fig.~\ref{Fig:Cv}, the various contributions to $C_V$ are illustrated.
When baryons, and quarks, become paired, as briefly described in
\S\ref{Sec:Pairing}, their contribution to $C_\mathrm{v}$ is strongly
suppressed at temperatures $T \ll T_\mathrm{c}$ ($T_\mathrm{c}$ being the corresponding
critical temperature).   Extensive baryon, and quark, pairing can thus
significantly reduce $C_\mathrm{v}$, but by at most a factor of order ten as
the leptons do not pair.
The crustal contribution is in principle
dominated by  neutrons in the inner crust but, as these are
certainly extensively paired, practically only the nuclear lattice and
electrons contribute. 

\subsection{Photon emission and the neutron star envelope}
\label{Sec:Envelope}

Thermal photons from the neutron star surface are emitted at the photosphere, which is usually in an
atmosphere but could be a solid surface in the presence of a very strong magnetic field.
The atmosphere, which is only a few centimeters thick, presents a temperature gradient. Consequently,  photons of
increasing energy escape from deeper and hotter layers.
It is customary to define an {\em effective} temperature, $T_e$, so that the total surface photon luminosity,
by analogy with the blackbody emission, is written as
\be
L_\gamma = 4 \pi R^2 \sigma_\mathrm{SB} T_e^4 \,,
\label{Eq:Lgamma1}
\ee
where $\sigma_\mathrm{SB}$ is the Stefan-Boltzmann constant.
Observationally, $L_\gamma$ and $T_e$ are red-shifted and Eq.~(\ref{Eq:Lgamma1}) is rewritten as
\be
L_{\gamma \, \infty}= 4 \pi R_\infty^2 \sigma_\mathrm{SB} T_{e \, \infty}^4
\label{Eq:Lgamma2}
\ee
where $L_{\gamma \, \infty} = \mathrm{e}^{2\phi} L_\gamma$, $T_{e \, \infty} = \mathrm{e}^{\phi} T_e$,
and $R_\infty = \mathrm{e}^{-\phi} R$.
Here $\mathrm{e}^{-\phi} = 1+z$, with $z$ being the red-shift, and $\mathrm{e}^{2\phi}$ is the $g_{00}$
coefficient of the Schwarzschild metric, i.e.,
\be
\mathrm{e}^\phi \equiv \sqrt{1-\frac{2GM}{Rc^2}} \,.
\label{Eq:redshift}
\ee
Notice that $R_\infty$ has the physical interpretation of being the star's radius that an observer would
measure trigonometrically, if this were possible.

In a detailed cooling calculation, the time evolution of the whole temperature profile in the star
is followed. 
However, the uppermost layers have a thermal time-scale much shorter than the interior of
the star and are practically always in a steady state.
It is hence common to treat these layers separately as an {\em envelope}.
Encompassing a density range from $\rho_b$ at its bottom (typically $\rho_b = 10^{10}$ g cm$^{-3}$)
up to that at the photosphere, and a temperature range from $T_b$ to $T_e$, the envelope is about
one hundred meters thick.
Due to the high thermal conductivity of degenerate matter, stars older than a few decades have 
an almost uniform internal temperature except within the envelope which  acts as a thermal
blanket insulating the hot interior from the colder surface.
A simple relationship between $T_b$ and $T_e$ can be formulated as in \cite{Gudmundsson:1983kx}:
\be
T_e \simeq 10^6 \left(\frac{T_b}{10^8 \, \mathrm{K}}\right)^{0.5+\alpha}
\label{Eq:TbTe}
\ee
with $\alpha \ll 1$.
The precise $T_e - T_b$ relationship depends on the chemical composition of the envelope.
The presence of light elements, resulting in larger thermal conductivities, implies a larger $T_e$ for the same $T_b$
compared to the case of a heavy element envelope.
Magnetic fields also alter this $T_e - T_b$ relationship (see, e.g., \cite{Page:2006ly} for more details).

\subsection{Some simple analytical solutions}
\label{Sec:Analytical}

As the essential ingredients entering Eq.~(\ref{Eq:Cooling}) can all be approximated by power-law functions,
one can obtain simple and illustrative analytical solutions.
Let us write 
\be
C_\mathrm{V} = C_9 \cdot T_9 \, ,
\;\;\;\;\;\;\;\;\;\;
L_\nu = N_9 \cdot T_9^8 \,  \qquad {\rm and} \qquad
\;\;\;\;\;\;\;\;\;\;
L_\gamma = S_9 \cdot T_9^{2+4\alpha} \, ,
\label{Eq:PL1}
\ee
where $T_9 \equiv T/(10^9 \; \mathrm{K})$. As written,
$L_\nu$ considers slow neutrino emission involving 
five degenerate fermions from the modified Urca and
the similar bremsstrahlung processes, summarized in Table~\ref{Tab:Nu}. The photon luminosity
$L_\gamma$ is obtained from Eq.~(\ref{Eq:Lgamma1}) using the simple expression in Eq.~(\ref{Eq:TbTe}).
Typical values are $C_9 \simeq 10^{39}$ erg K$^{-1}$, $N_9 \simeq 10^{40}$ erg s$^{-1}$
and $S_9 \simeq 10^{33}$ erg s$^{-1}$
(see Table 3 in \cite{Page:2006ly} for more details).
In young stars neutrinos dominate the energy losses (in the so-called {\em neutrino cooling era}) and photons
take over after about $10^5$ yrs (in the {\em photon cooling era}).

\noindent
{\bf Neutrino cooling era:} In this case, 
we can neglect $L_\gamma$ in Eq.~(\ref{Eq:Cooling}) and find
\be
t = \frac{10^9 C_9}{6N_9} \left(\frac{1}{T_9^6}-\frac{1}{T_{0 \, 9}^6}\right)
\;\;\;\;\;\;
\longrightarrow
\;\;\;\;\;\;
T = 10^9 \, \mathrm{K} (\tau_\mathrm{MU}/t)^{1/6}
\;\;\; (\mathrm{when} \;T \ll T_0)
\label{Eq:Simple_nu}
\ee
with a MU cooling time-scale $\tau_\mathrm{MU} = 19^9 C_9/6N_9 \sim 1$ yr when
the star reaches the asymptotic solution ($T \ll T_0$, $T_0$ being the initial temperature at
time $t=t_0 = 0$).

\noindent
{\bf Photon cooling era:} In this era,  
$L_\nu$ becomes negligible compared to $L_\gamma$ so that we obtain
\be
t = t_1 + \frac{C_9}{4\alpha S_9 10^{9+36\alpha}} 
        \left(\frac{1}{T_9^{4\alpha}}-\frac{1}{T_{1 \, 9}^{4\alpha}}\right)
\;\;\;\;\;\;
\longrightarrow
\;\;\;\;\;\;
T \propto t^{-\frac{1}{4\alpha}}
\;\;\; (\mathrm{when} \;T \ll T_1)
\label{Eq:Simple_gamma}
\ee
where $T_1$ is $T$ at time $t_1$.
Notice here that the slope, $-1/4\alpha$, of the asymptotic solution is strongly affected by any small
change in the envelope structure, Eq.~(\ref{Eq:TbTe}).

\subsection{Some numerical solutions}
\label{Sec:Numerical}

\begin{figure*}[t]
\begin{center}
\includegraphics[width=.35\textwidth]{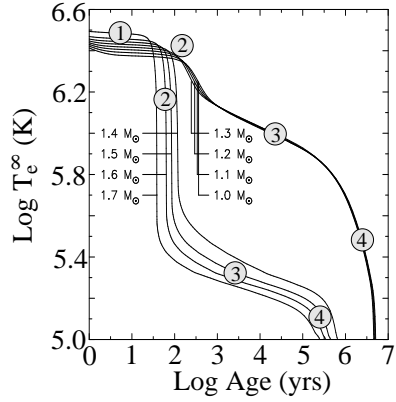}
\end{center}
\caption{Slow versus fast cooling (see text for details of stages 1 through 4 marked in the figure).
Figure adapted from \cite{Page:1992nx}.}
\label{Fig:Cooling}
\end{figure*}

Numerical simulations of a cooling neutron star use an evolution code in which 
the energy balance and energy transport equations 
in their fully general relativistic forms are solved, usually assuming spherical symmetry and
with a numerical radial grid of several hundred zones \footnote{Such a code, \texttt{NSCool}, is available at: \texttt{http://www.astroscu.unam.mx/neutrones/NSCool/}}.
A set of cooling curves that illustrate the difference between cooling driven by the modified Urca
and the direct Urca processes is presented in Fig.~\ref{Fig:Cooling}.
Cooling curves of eight different stars of increasing mass are shown, using an equation of state model,
from \cite{Prakash:1988oq}, which allows the DU process at densities above 
$1.25\times 10^{15}$ g cm$^{-3}$, i.e., above a critical neutron star mass of 1.35 $M_\odot$.
Notice that the equation of state used is a parametric one and its parameters were 
{\em specifically adjusted} to obtain a critical mass of 1.35 $M_\odot$ which falls within
the expected range of isolated neutron star masses; other equations of state can
result in very different critical masses.
The difference arising from  {\em slow} and {\em fast} neutrino processes is clear.

Four successive cooling stages are marked in the figure.
The {\em neutrino cooling era} is marked stage 3 and the {\em photon cooling era} is marked stage 4.
As the figure shows $T_e$, and since $T_e \sim T^{0.5}$, the slopes of the cooling curves in this figure
are rescaled from the values of Eq.~(\ref{Eq:Simple_nu}) and (\ref{Eq:Simple_gamma}), i.e., they become
$\approx -1/12$ and $\approx -1/8\alpha$, respectively.
The simple analytical model of the previous section considered a single temperature in the stellar interior.
For a young star, whose age is smaller that its thermal relaxation time, the interior temperature
has a very complicated radial profile. 
In particular, the core cools much faster than the crust, due to its much stronger neutrino emission.
During stage 1  in Fig.~\ref{Fig:Cooling}, the surface temperature is controlled by processes
occurring in the outer layers of the crust and is totally independent of the temperature deeper in the star.
(One can appreciate that the more massive stars have a larger $T_e$ at that time; this is mostly due to the fact that their blanketing envelopes are thinner than in low mass stars.)
The rapid drop in $T_e$ occurring during stage 2 corresponds to the thermal relaxation of the crust; 
the star's age becomes comparable to the heat transport time-scale from the crust to the core and the
crustal heat flows into the core.
After stage 2, the stellar interior is essentially isothermal with a strong temperature gradient
present only in the envelope and the simple analytical solutions presented above apply.

\section{Pairing and its Effects}
\label{Sec:Pairing}

\begin{figure*}[t]
\begin{center}
\includegraphics[width=.50\textwidth]{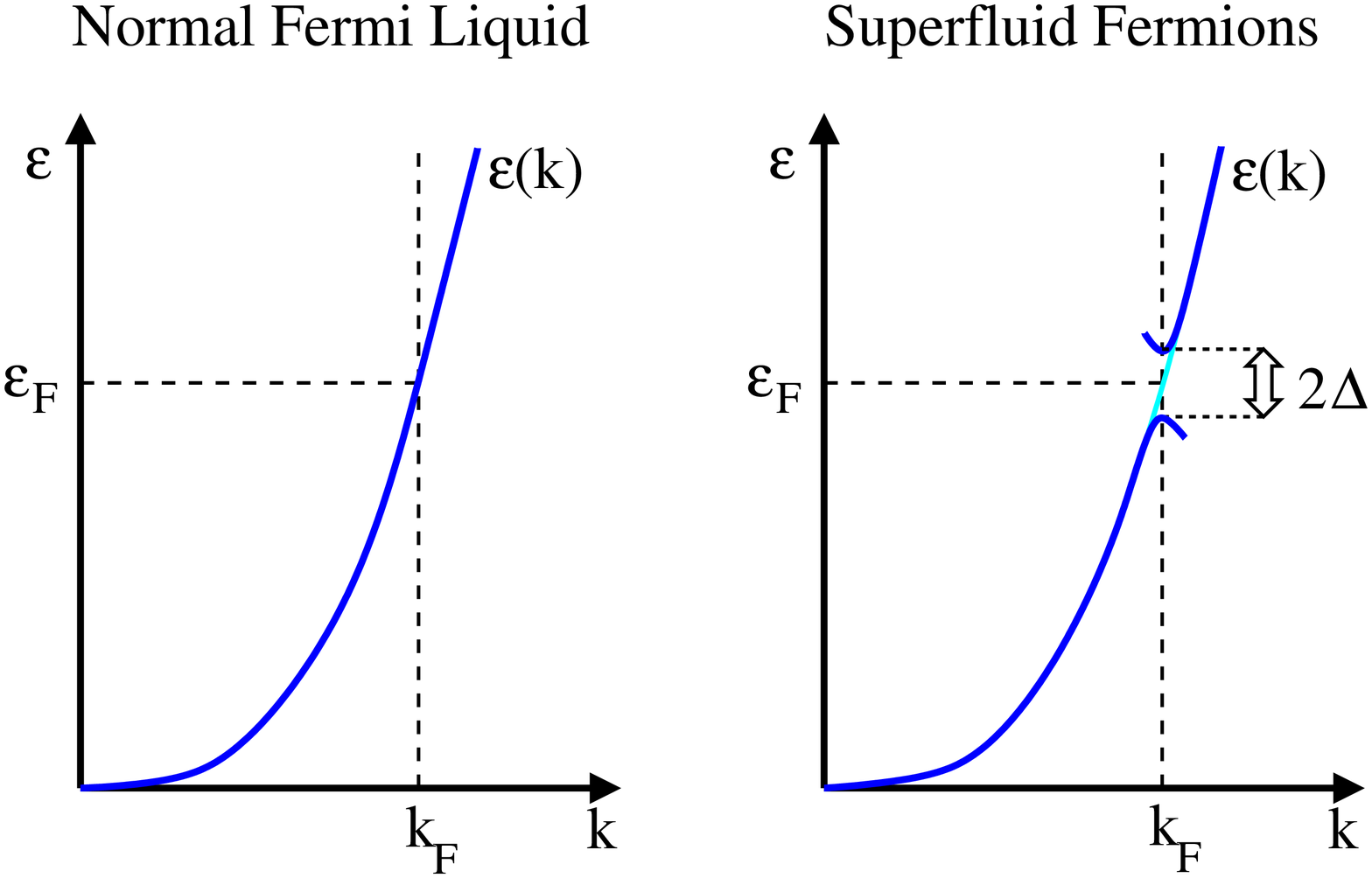}
\end{center}
\caption{Comparison of quasi-particle spectra, $\epsilon(k)$, i.e., the pole in the
2-point Green's function, for a normal and a superfluid Fermi liquid.
The reorganization of particles at $\epsilon \sim \epsilon_F$ into Cooper pairs results
in the development a gap $2 \Delta(k)$ in the spectrum so that no particle can have
an energy between $\epsilon_F - \Delta(k_F)$ and $\epsilon_F + \Delta(k_F)$.}
\label{Fig:SP_Spectrum}
\end{figure*}

\begin{figure*}[b]
\begin{center}
\includegraphics[width=.95\textwidth]{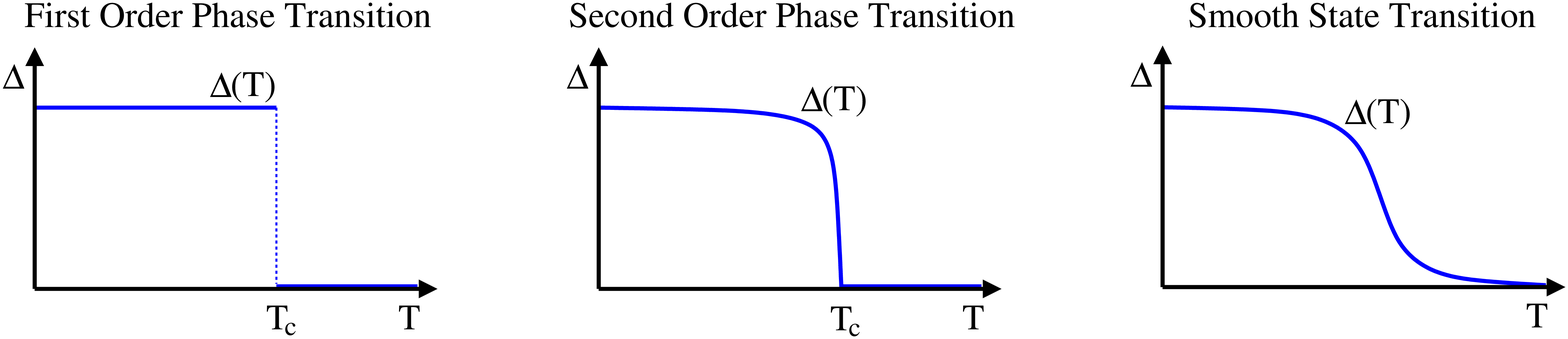}
\end{center}
\caption{
Temperature evolution of the state of a system parametrized by an "order" parameter, $\Delta(T)$.
\newline
{\bf First order phase transition:} 
discontinuous change of $\Delta$ at $T_c$;
latent heat due to the entropy difference between the two states.
(Examples: solid $\leftrightarrow$ liquid; liquid $\leftrightarrow$ gas below the critical point.)
\newline
{\bf Second order phase transition:}
continuous change of $\Delta$ but with a discontinuity in the slope at $T_c$;
no latent heat but a discontinuity in specific heat.
(Examples: superfluid $\leftrightarrow$ normal fluid; ferromagnetic $\leftrightarrow$ paramagnetic.)
\newline
{\bf Smooth state transition:}
continuous change of $\Delta$ with no critical temperature;
increase of specific heat in the regime where $\Delta(T)$ changes rapidly.
(Examples: liquid $\leftrightarrow$ gas above the critical point; atomic gas $\leftrightarrow$ plasma.)
}
\label{Fig:Order}
\end{figure*}

Pairing, which induces superfluidity in the case of degenerate neutral fermions and superconductivity
for charged fermions, is expected to occur between neutrons/protons in the interior of
neutron stars \cite{Migdal:1959bh}.
The {\em Cooper Theorem} \cite{Cooper:1956qf} states that, in a system of degenerate fermions 
the Fermi surface is unstable, at $T=0$, due to the formation of Cooper pairs if there is an attractive interaction in some spin-angular momentum channel. 
The essence of the BCS theory \cite{Bardeen:1957dq} is that as a result of this instability there is a
collective reorganization of the particles at energies around the Fermi energy and the appearance of a gap
in the quasi-particle spectrum (see Fig.~\ref{Fig:SP_Spectrum}) which is the binding energy of a Cooper pair.
At high enough temperature the gap $\Delta(T)$ vanishes and the system is in the normal state.
The transition to the superfluid/superconducting state is a second order phase transition and the gap
$\Delta(T)$ is its order parameter (see Fig.~\ref{Fig:Order}). Explicitly, 
$\Delta(T)=0$ when $T > T_c$ and, when $T$ drops below $T_c$, $\Delta(T)$ grows rapidly but continuously,
with a discontinuity in its slope at $T=T_c$.
This is in sharp contrast with a first order phase transition, in which the transition occurs
entirely at $T_c$ (see left panel of Fig.~\ref{Fig:Order}) and will be of paramount importance for our purpose.
In the BCS theory, which remains approximately valid for nucleons,
the relationship between the gap and $T_c$ is
\be
\Delta(T=0) \simeq 1.75 k_B T_c \,.
\label{Eq:DTc}
\ee

In a normal Fermi system at $T=0$, all particles are in states with energies $\epsilon \le \epsilon_F$.
When $T >0$, states with energies $\epsilon \simgreater \epsilon_F$ can be smoothly occupied
(left panel of Fig.~\ref{Fig:SP_Spectrum}) resulting
in a smearing of the particle distribution around $\epsilon_F$ in a range $\sim k_BT$.
It is precisely this smooth smearing of energies around $\epsilon_F$ which produces the linear $T$
dependence of $c_v$, \S~\ref{Sec:Cv}, and the $T^6$ or $T^8$ dependence of the neutrino
emissivities, \S~\ref{Sec:nuT}.

\begin{figure*}[t]
\begin{center}
\includegraphics[width=.70\textwidth]{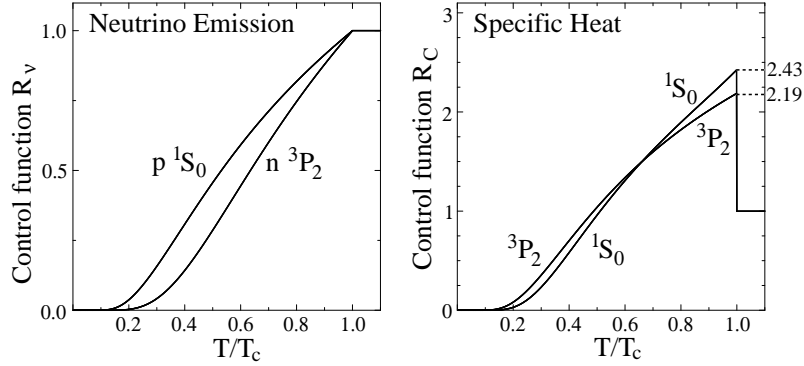}
\end{center}
\caption{Control functions for MU neutrino emission (left panel) and specific heat (right panel).}
\label{Fig:Control}
\end{figure*}

In a superfluid/superconducting Fermi system at $T=0$, all particles are in states with energies 
$\epsilon \le \epsilon_F$ (actually, $\epsilon \le \epsilon_F - \Delta$).
When $0< T < T_c$, states with energy $\epsilon > \epsilon_F$ 
(actually, $\epsilon \ge \epsilon_F + \Delta$) can be populated.
However, in contrast to the smooth filling of levels above $\epsilon_F$ in the case of a normal Fermi liquid,
the presence of the $2 \Delta(T)$ gap in the spectrum implies that
the occupation probability is strongly suppressed by a Boltzmann factor $\sim \exp[-2 \Delta(T)/k_BT]$.
Consequently, all physical properties/processes depending on thermally excited particles, such
as the specific heat and the neutrino emission processes described in \S~\ref{Sec:Neutrinos},
are strongly suppressed when $T\ll T_c$.
In practice, for numerical simulations of neutron star cooling, these suppressions are implemented
through "control functions" $R_*$ such that
\ba
c_\mathrm{v} \rightarrow c_\mathrm{v}^\mathrm{Paired} &=& R_c c_\mathrm{v}^\mathrm{Normal}
\label{Eq:Cvsupp}
\\
\epsilon_\nu \rightarrow \epsilon_\nu^\mathrm{Paired} &=& R_\nu \epsilon_\nu^\mathrm{Normal} \; .
\label{Eq:Nusupp}
\ea
There is a large family of such functions, for each process and they moreover depend on how many
of the participating particles are paired and on the specific kind of pairing; a few examples are
displayed in Fig.~\ref{Fig:Control} \cite{Yakovlev:2001dq}.

\subsection{Theoretical expectations on pairing gaps}
\label{Sec:Gaps}

\begin{figure*}[b]
\begin{center}
\includegraphics[width=.40\textwidth]{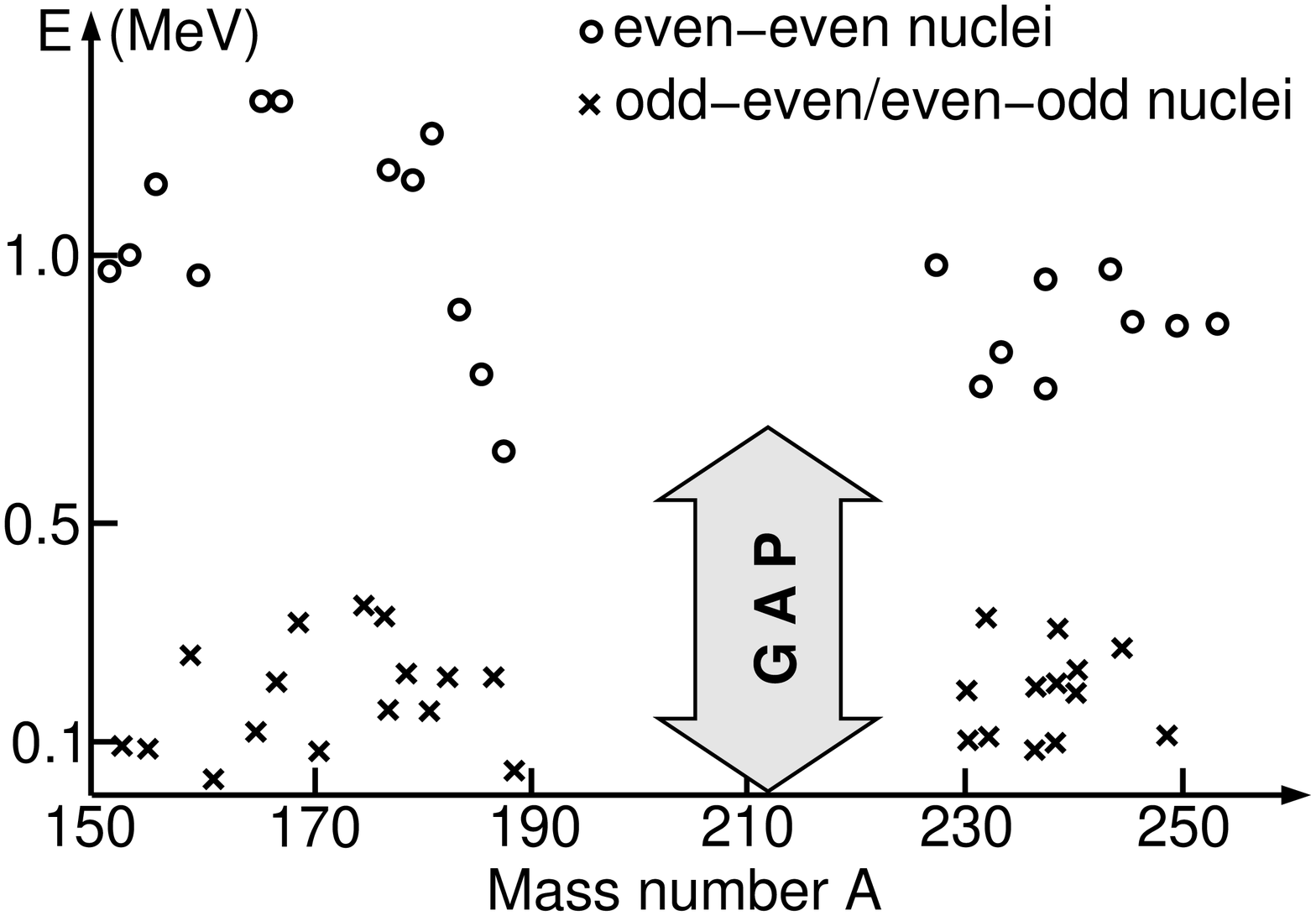}
\includegraphics[width=.45\textwidth]{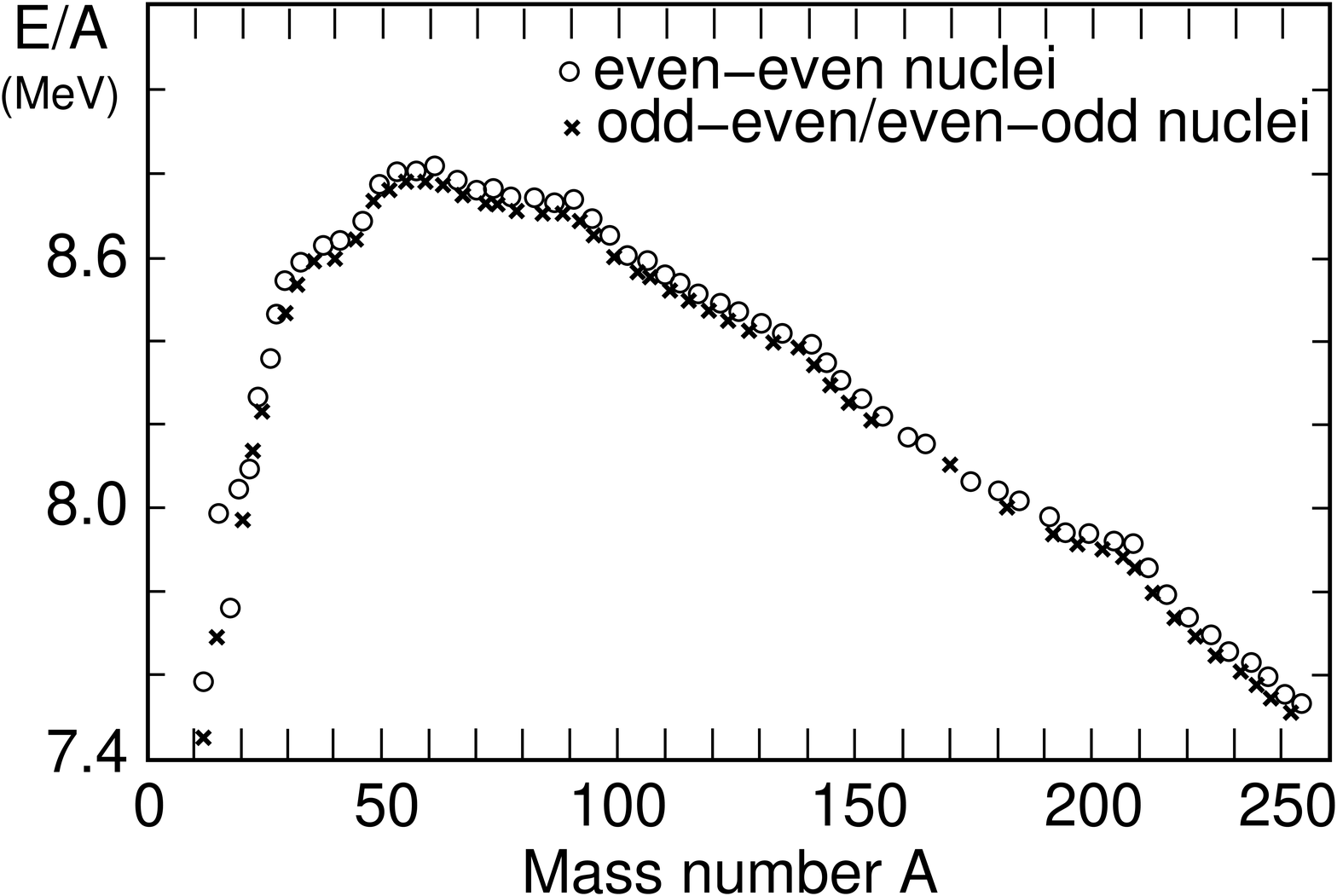}
\end{center}
\caption{Left panel: lowest excitation levels of nuclei (adapted from \cite{Bohr:1958fk}).
Right panel: binding energy per nucleon for the most beta-stable isobars.}
\label{Fig:Gap-Nuclei}
\end{figure*}

Soon after the development of the BCS theory, Bohr, Mottelson \& Pine \cite{Bohr:1958fk} 
pointed out 
that excitation energies of nuclei  show a gap, as shown in the left panel of Fig.~\ref{Fig:Gap-Nuclei}.
Even-even nuclei clearly require a finite minimal energy for excitation. 
This energy was interpreted as being the binding energy of the Cooper pair which must break to produce an excitation.
In contrast, odd nuclei do not show such a gap, and this is due to the fact that they have one
nucleon, neutron or proton, which is not paired and can be easily excited.
The right panel of Fig.~\ref{Fig:Gap-Nuclei} shows that pairing also manifests itself in the binding
energies, even-even nuclei being slightly more bound than odd nuclei\footnote{Notice that,
as a result of pairing, the only stable odd-odd nuclei are
$^2$H(1,1), $^6$Li(3,3), $^{10}$B(5,5), and $^{14}$N(7,7). All heavier odd-odd nuclei are beta-unstable
and decay into an even-even nucleus.}.

\begin{figure*}[t]
\begin{center}
\includegraphics[width=.40\textwidth]{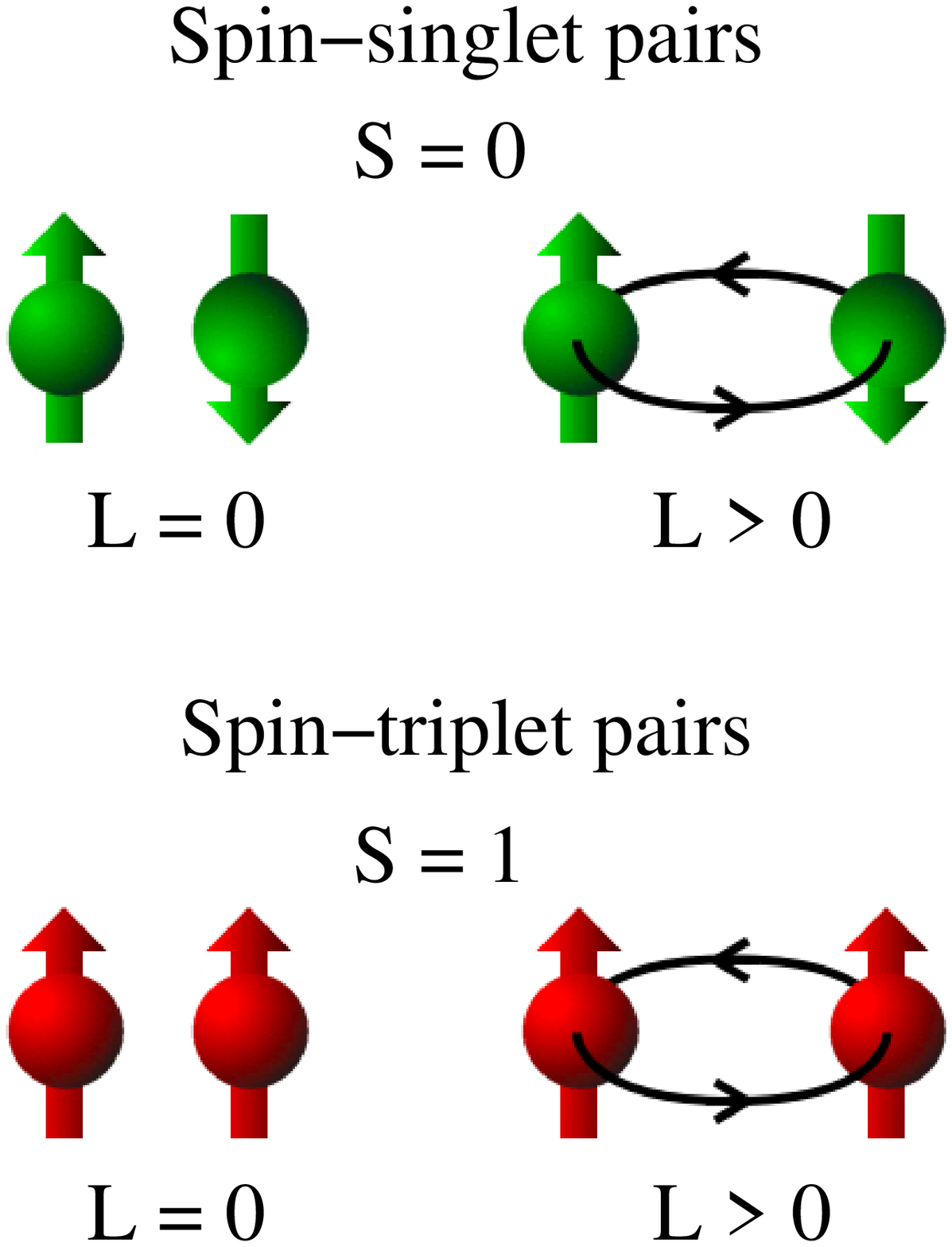}
\includegraphics[width=.40\textwidth]{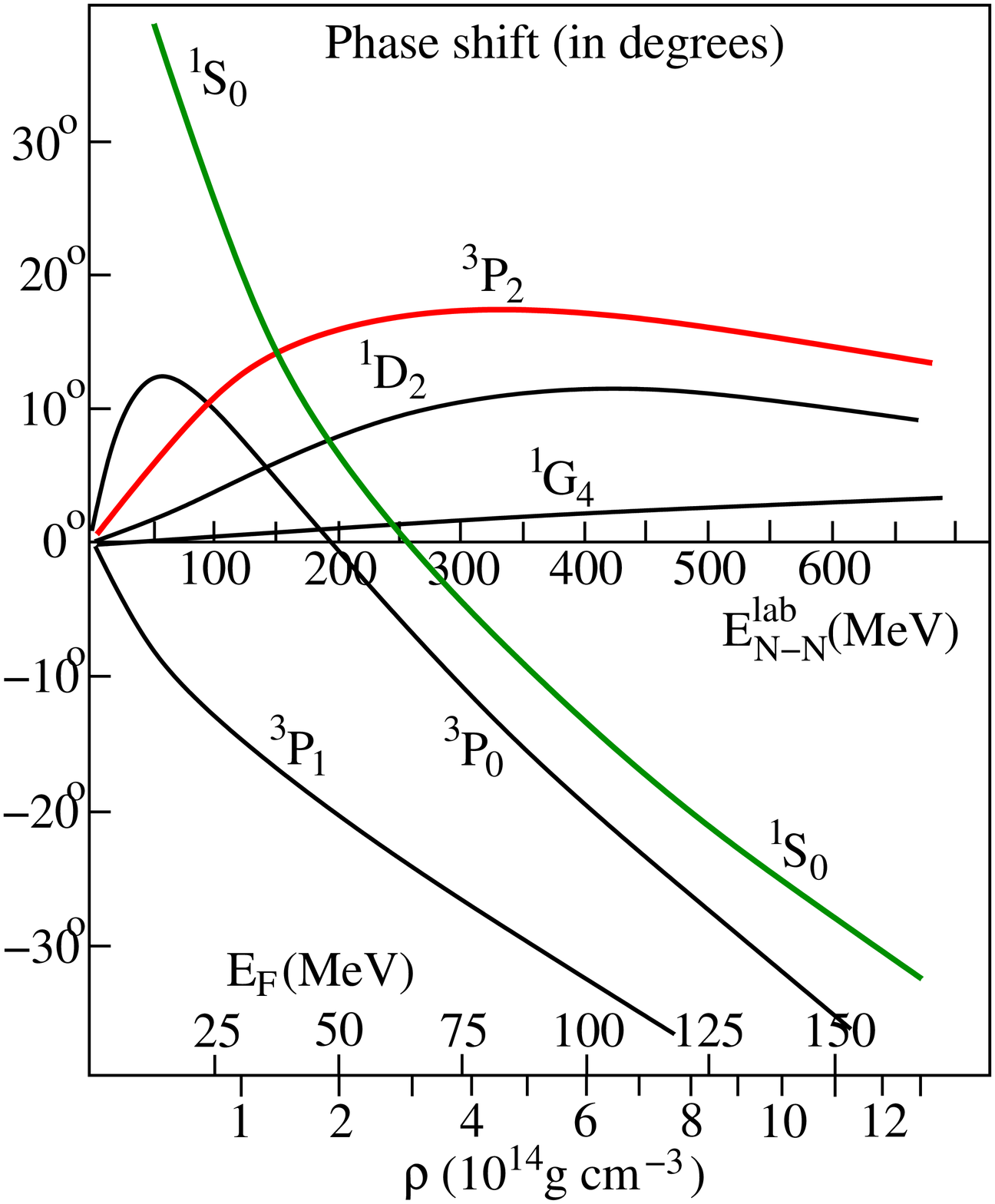}
\end{center}
\caption{
Left panel: 
possible spin-angular momentum combinations for Cooper-pairs.
Right panel:
phase shifts for N-N scattering as a function of the laboratory energy (middle axis)
or the neutron Fermi energy and density for a neutron star interior (lower axis).
Adapted from \cite{Tamagaki:1970uq}.}
\label{Fig:CooperPairs}
\end{figure*}

As a two-particle bound state, the Cooper pair can appear in many spin-orbital angular momentum
states (see left panel of Fig.~\ref{Fig:CooperPairs}).
In terrestrial superconducting metals, the Cooper pairs are generally in the $^1$S$_0$ channel, i.e., 
spin-singlets with $L=0$ orbital angular momentum, whereas in liquid $^3$He they are in spin-triplet states.
What can we expect in a neutron star ?
In the right panel of Fig.~\ref{Fig:CooperPairs},  we adapt a figure from one of the first works to study
neutron pairing in the neutron star core \cite{Tamagaki:1970uq} showing laboratory measured
phase-shifts from N-N scattering.  A positive phase-shift implies an attractive interaction.
From  this figure, one can expect that nucleons could pair in a spin-singlet state, $^1$S$_0$, at low densities,
whereas a spin-triplet, $^3$P$_2$, pairing should appear at higher densities.
We emphasize that this is only a {\em presumption} as medium effects can  
strongly affect particle interactions.

A simple model can illustrate the difficulty in calculating pairing gaps.
Consider a dilute Fermi gas with a weak, attractive, interaction potential $U$.
The interaction is then entirely described by the corresponding scattering length, 
$a$, \footnote{The scattering length $a$ is related to $U$ by $a = (m/4\pi \hbar) U_0$ with 
$U_\mathbf{k} = \int d^3r \, \exp(i \mathbf{k}\cdot \mathbf{r}) \, U(\mathbf{r}) $.}
which is
negative for an attractive potential.
In this case, the model has a single dimensionless parameter, $|a|k_F$, and the dilute gas corresponds to $|a|k_F \ll 1$.
Assuming the pairing interaction is just the bare interaction $U$ (which is, improperly, called the
{\em BCS approximation}), the gap equation at $T=0$  can be solved analytically, giving the
{\em weak-coupling BCS-approximation} gap:
\be
\Delta(k_F)  \;\; \longrightarrow \;\; \Delta_{BCS}(k_F) =
\frac{8}{e^2} \left(\frac{\hbar^2 k_F^2}{2M}\right) \exp \left[ -\frac{\pi}{2|a|k_F} \right]
\;\;\;\;\;\; \mathrm{when} \;\;\;\;\;\;
|a| k_F \rightarrow 0 \; .
\label{Eq:Delta_BCS}
\ee
This result is bad news: the gap depends exponentially on the pairing potential $U$.
The Cooper pairs have a size of the order of 
$\xi \sim \hbar v_F/\Delta$ (the {\em coherence length}) and thus
$\xi k_F \sim \exp [ \mathbf{+} \pi/2|a|k_F] \gg 1$ in the weak coupling limit.
There appear to be many other particles within the pair's coherence length.
These particles will react, and can screen or un-screen, the interaction.
Including this medium polarization on the pairing is called {\em beyond BCS}, and in the weak
coupling limit its effect has been calculated analyticaly \cite{Gorkov:1961kx}, giving
\be
\Delta(k_F)  \;\; \longrightarrow \;\; \Delta_{GMB}(k_F) = 
\frac{1}{(4\mathrm{e})^{1/3}} \Delta_{BCS}(k_F) 
\simeq 0.45 \Delta_{BCS}(k_F)
\;\;\;\;\;\; \mathrm{when} \;\;\;\;\;\;
|a| k_F \rightarrow 0 \; .
\label{Eq:Delta_GMB}
\ee
So, screening by the medium reduces the gap by more than a factor two, even in an extremely
dilute system.

\begin{figure*}[t]
\begin{center}
\includegraphics[width=.99\textwidth]{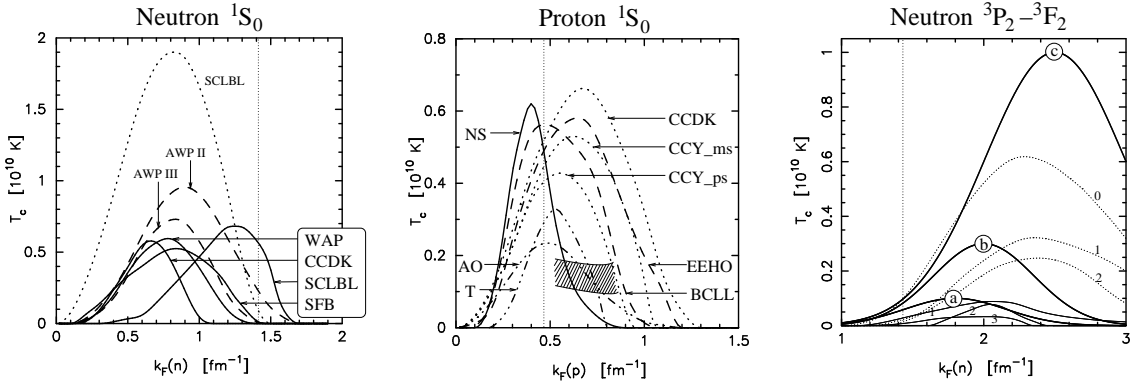}
\end{center}
\caption{Three collections of predicted pairing critical temperatures for neutrons in the
$^1$S$_0$ and $^3$P$_2$ channels and protons in the $^1$S$_0$ channel.
See \cite{Page:2004zr} for references to the original works.}
\label{Fig:3Gaps}
\end{figure*}

Pairing correlations in nuclei are part of everyday nuclear physics,
and a significant amount of work has also been devoted to the neutron star environment
(see, e.g., \cite{Dean:2003vn} and \cite{Lombardo:2001ys} for reviews).
We show in Fig.~\ref{Fig:3Gaps} three sets of predicted $T_c$ for the neutron star interior.
At low density, corresponding to the neutron star crust in the regime of dripped neutrons,
the expectation of a neutron $^1$S$_0$ superfluid is amply confirmed by the models.
This regime was also illustrated in the inset A of Fig.~\ref{Fig:NStar}.
At higher densities, corresponding to the neutron star core, the situation is much more
ambiguous.
Due to their low concentrations, protons have small Fermi momenta in the core and
are expected to form a $^1$S$_0$ superconductor.
There is, however, a significant uncertainty in the size of their gap, with predicted values of $T_c$ ranging
from $\sim 10^9$ K up to $6\times 10^9$ K, and a larger uncertainty in the range of Fermi
momenta in which $\Delta(k_F)$ is non-zero, which translates into an uncertainty of a factor
of more than 3 on the density range covered by the superconductor.
In the "pessimistic" case protons would be superconducting only in the outer part of the core,
whereas in the "optimistic" case the whole core may be superconducting.

For neutrons in the neutron star core, there is no agreement between the many published models
about either the maximum value of $T_c$ or on the density range in which pairing is significant.
Notice that, due to the tensor interaction, pairing is expected to be in the
$^3$P$_2$-$^3$F$_2$ channel.
However, even the best models of the N-N interaction {\em in vacuum} fail to reproduce the
measured phase shift in the $^3$P$_2$ channel \cite{Baldo:1998zr}.
 Due to medium polarization
a long-wavelength tensor force appears  that is not
present in the {\em in vacuum} interaction and results in a strong suppression of the gap
\cite{Schwenk:2004ly}.
Recently, the effect of three-body forces (TBF), absent in the laboratory N-N scattering experiment,
has been considered.
TBF are necessary to reproduce the nuclear saturation density; they are, in the bulk, repulsive and
their importance grows with increasing density.
However, it was found in \cite{Zhou:2004ve,2008PhRvC..78a5805Z} that, at the Fermi surface, they are
strongly attractive in the $^3$P$_2$-$^3$F$_2$ channel and result in very large neutron 
$^3$P$_2$-$^3$F$_2$ gaps.
Other delicate issues are the effect of the proton contaminant and the likely development of a
$\pi^0$ condensate\footnote{In the presence of a charged $\pi^-$ condensate a new Urca neutrino
emission pathway is open, see Table~\ref{Tab:Nu} and Eq~(\ref{Eq:pionDU}). 
The development of a neutral $\pi^0$ condensate has, however, little effect on neutrino emission.}
which also strongly affects the size of the neutron  (and proton) gap(s).
In short, the size and extent in density of the neutron $^3$P$_2$-$^3$F$_2$ gap in the neutron star
core is poorly known.

\subsection{The Cooper pair neutrino process}
\label{Sec:PBF}

\begin{figure*}[b]
\begin{center}
\includegraphics[width=.60\textwidth]{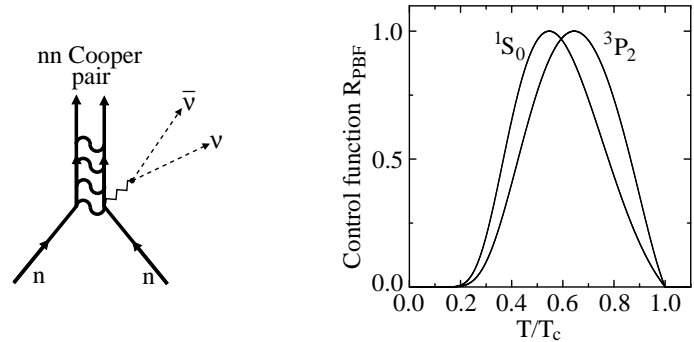}
\end{center}
\caption{
Left panel: Feynman diagram for $\nu-\overline{\nu}$ emission from the pair breaking and formation (PBF) process. 
Right panel: control functions $R_\mathrm{PBF}$ for the PBF process.
}
\label{Fig:PBF}
\end{figure*}

The formation of the fermonic pair condensate also triggers a new neutrino emission process,
which has been termed as the "pair breaking and formation", or PBF, process \cite{Schaab:1997qf}.
When an ff pair (f = n, p, or any fermion undergoing pair condensation) forms, its binding energy
can be emitted as a $\nu-\overline{\nu}$ pair.
Under the right conditions, this PBF process can be the dominant cooling agent in the evolution 
of a neutron star \cite{Page:1998bh}.
Such efficiency is due to the fact that the pairing phase transition is second order in nature.
During the cooling of the star, the phase transition starts when the temperature $T$ reaches $T_c$ when
pairs begin to form, but thermal agitation will constantly induce the breaking of pairs
with subsequent re-formation and possible neutrino pair emission.

The emissivity of the PBF process can be written as
\ba
\epsilon^\mathrm{PBF} = \frac{12 G_F^2 m_{f}^{*} p_{F,f}}{15 \pi^5 \hbar^{10} c^6}
\left(k_B T\right)^{7}  a_{f,j} R_j\left[\Delta_j(T)/T\right] 
\nonumber
\\
= 3.51\times 10^{21}~
\frac{\mathrm{erg}}{\mathrm{cm}^3~\mathrm{s}} \times
\tilde{m}_f \, \tilde{p}_{F,f}
\, T_9^7 \, a_{f,j} \;R_j\left[\Delta_j(T)/T\right]
\label{Eq:Q_PBF}
\ea
for a fermion $f$ in a pairing state $j = ^1$S$_0$ or $^3$P$_2$.
The coefficients $a_{f,j}$ depend on the type of fermion and on vector and axial couplings $C_V$ and $C_A$
(see, e.g., \cite{Page:2009ly}).
The control functions $R_j$ are plotted in the right panel of Fig.~\ref{Fig:PBF}. These functions
encapsulate the effect that the PBF process turns on when $T$ reaches $T_c$
and practically turns off at $T \simless 0.2 T_c$ when there is not enough thermal energy to break pairs.
The PBF process was first discovered by Flowers, Ruderman, and Sutherland \cite{Flowers:1976vn} and,
independently, by Voskresensky and Senatorov \cite{Voskresensky:1987uq}.
It was, however, completely overlooked for 20 years (?) until finally implemented in a cooling
calculation in \cite{Schaab:1997qf} and its importance emphasized in \cite{Page:1998bh}.

An alternative way of looking at the PBF process  is simply as an
inter band transition of a nucleon \cite{Yakovlev:1999cr}.
Considering the particle spectrum in a paired state, right panel of Fig.~\ref{Fig:SP_Spectrum},
the lower branch (with $\epsilon < \epsilon_F - \Delta$) corresponds to paired particles whereas
the upper branch to excited ones, i.e., coming from a "broken pair" which hence left a hole in the
lower branch.
A transition of a particle from the upper branch to a hole in the lower branch corresponds to the
formation of a Cooper pair.

\section{Cooling of Superfluid Neutron Stars}
\label{Sec:Cooling_Sf}

\begin{figure*}[t]
\begin{center}
\includegraphics[width=.38\textwidth]{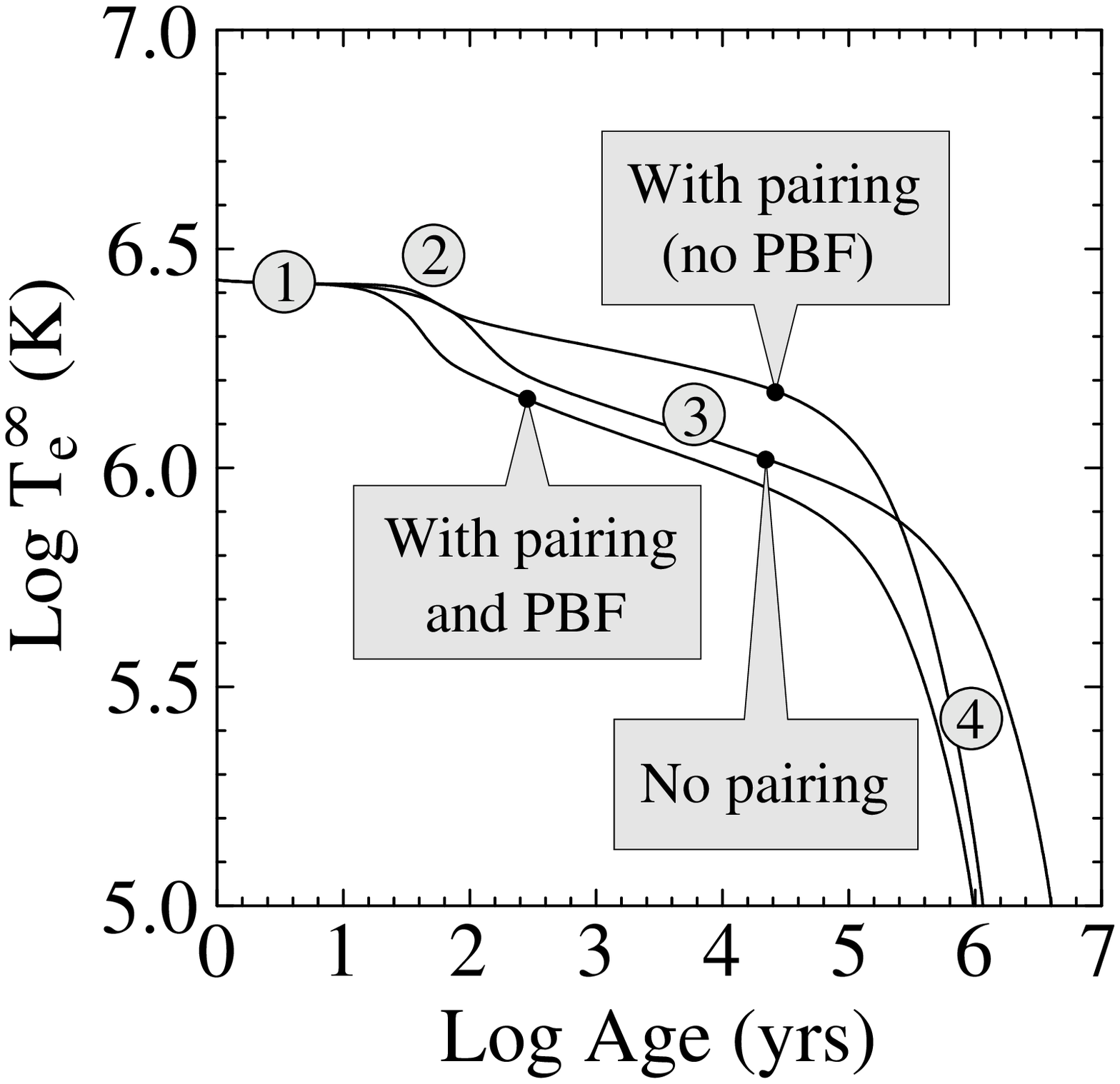}
\hspace{0.5cm}
\includegraphics[width=.53\textwidth]{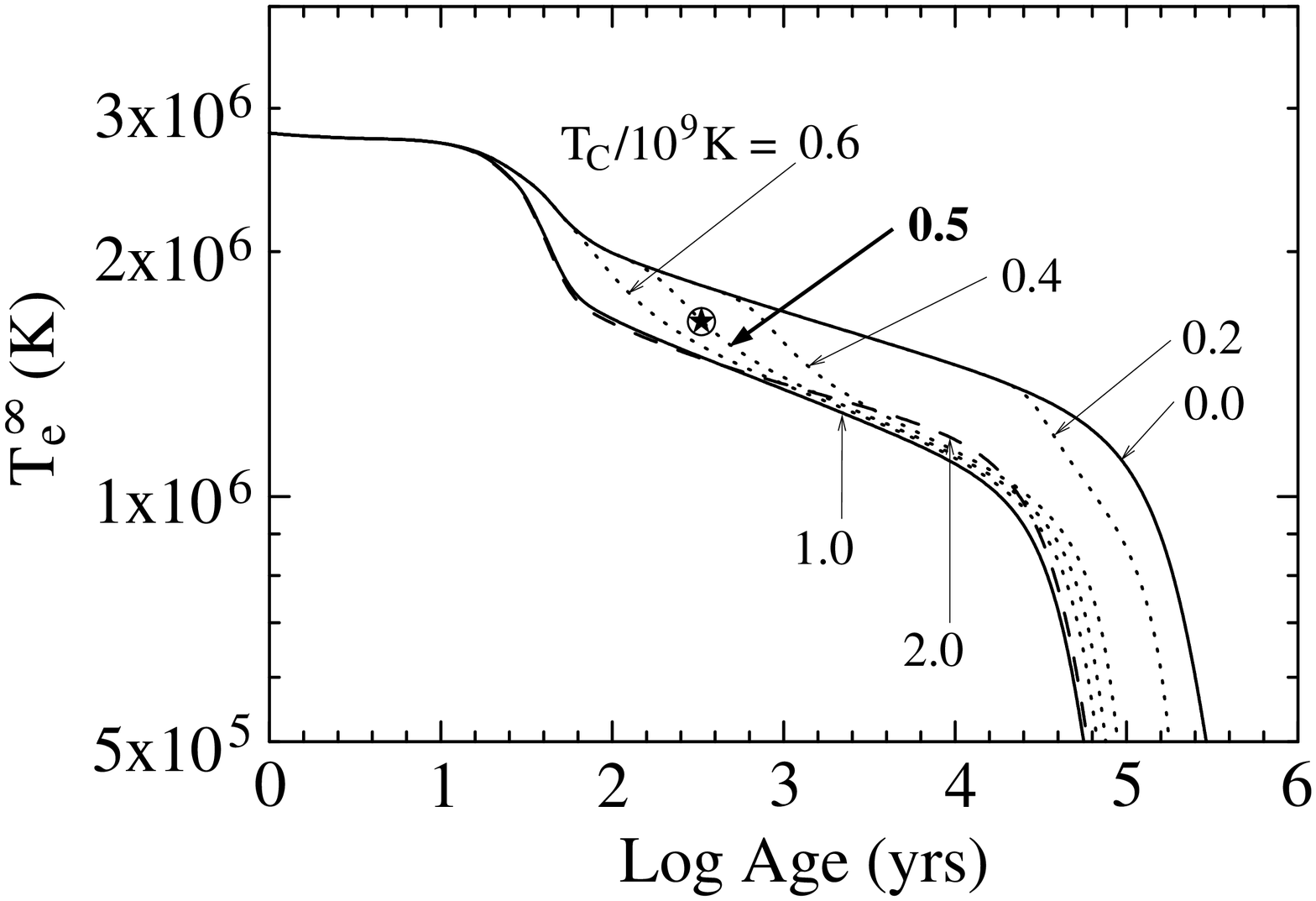}
\end{center}
\caption{Left panel: three models of cooling neutron stars illustrating the impact of pairing.
Right panel: exploration of the effect of a late onset of neutron superfluidity \cite{Page:2009ly,Page:2011dz}.
The curves are labeled according to the maximum value of the neutron $^3$P$_2$ $T_c$:
$T_C = \mathrm{max} \; T_c(k_F)$.
No proton superconductivity is taken into account in these models. 
The star marks the observed value of $T_e^\infty$ for the Cas A neutron star at an age of 330 yrs.
See text for description.}
\label{Fig:Cooling_SF}
\end{figure*}

In order to complement the brief description of cooling presented in \S~\ref{Sec:Numerical}, 
we now include the effects of pairing, but restrict ourself 
to cases in which no fast neutrino emission process is allowed
(the {\em minimal cooling} scenario \cite{Page:2004zr,Page:2009ly}).
In the left panel of Fig.~\ref{Fig:Cooling_SF},
three examples of the thermal evolution of a $1.4 \; M_\odot$ star are shown.  
The 4 cooling stages, introduced in Fig.~\ref{Fig:Cooling}, are similarly marked in this figure.
The first model does not include any effect from pairing, and is very similar to the $1.4 \; M_\odot$ model
of Fig.~\ref{Fig:Cooling} (but not completely identical because of several improvements in the microphysics
with respect to the older calculations of \cite{Page:1992nx}).
The other two models include nucleon pairing and in both cases the suppressing effects on the neutrino
emission as well as the modification of the specific heat are taken into account (see 
Eqs.~(\ref{Eq:Nusupp}) and (\ref{Eq:Cvsupp})), but the PBF process is artificially turned off in one case.
When Cooper pair neutrino emission from the PBF process is turned off,  one can clearly see
the effect of the suppression of the modified Urca and bremsstrahlung processes during stage 3 which 
results in a much warmer star. One can also see  
the effect of the suppression of the specific heat which results in a faster
cooling during the photon cooling era (stage 4).
During stage 3, the specific heat is also suppressed, but its effect is not as pronounced as in
stage 4.
In the presence of the PBF process, however, the star is significantly colder during stage 3, the PBF process being more efficient than the MU process in the unsuppressed case (i.e., no pairing).
In stage 4, the two trajectories with pairing, but  with and without the PBF process, converge as 
cooling is now driven by the photons and only the specific heat suppression by pairing is relevant. 
Notice that during stage 2, the crust relaxation period, the model with pairing (and PBF) has a shorter relaxation
time because of the suppression of the neutron specific heat from the $^1$S$_0$ gap in the inner crust
(the same effect is present in the model with pairing and no PBF, but is barely visible because the core is too hot).
In stage 1, the three models give identical $T_e$'s as at that age $T_e$ is controlled by the
evolution of the upper layers where no superfluid is present.

\begin{figure*}[t]
\begin{center}
\includegraphics[height=.20\textheight]{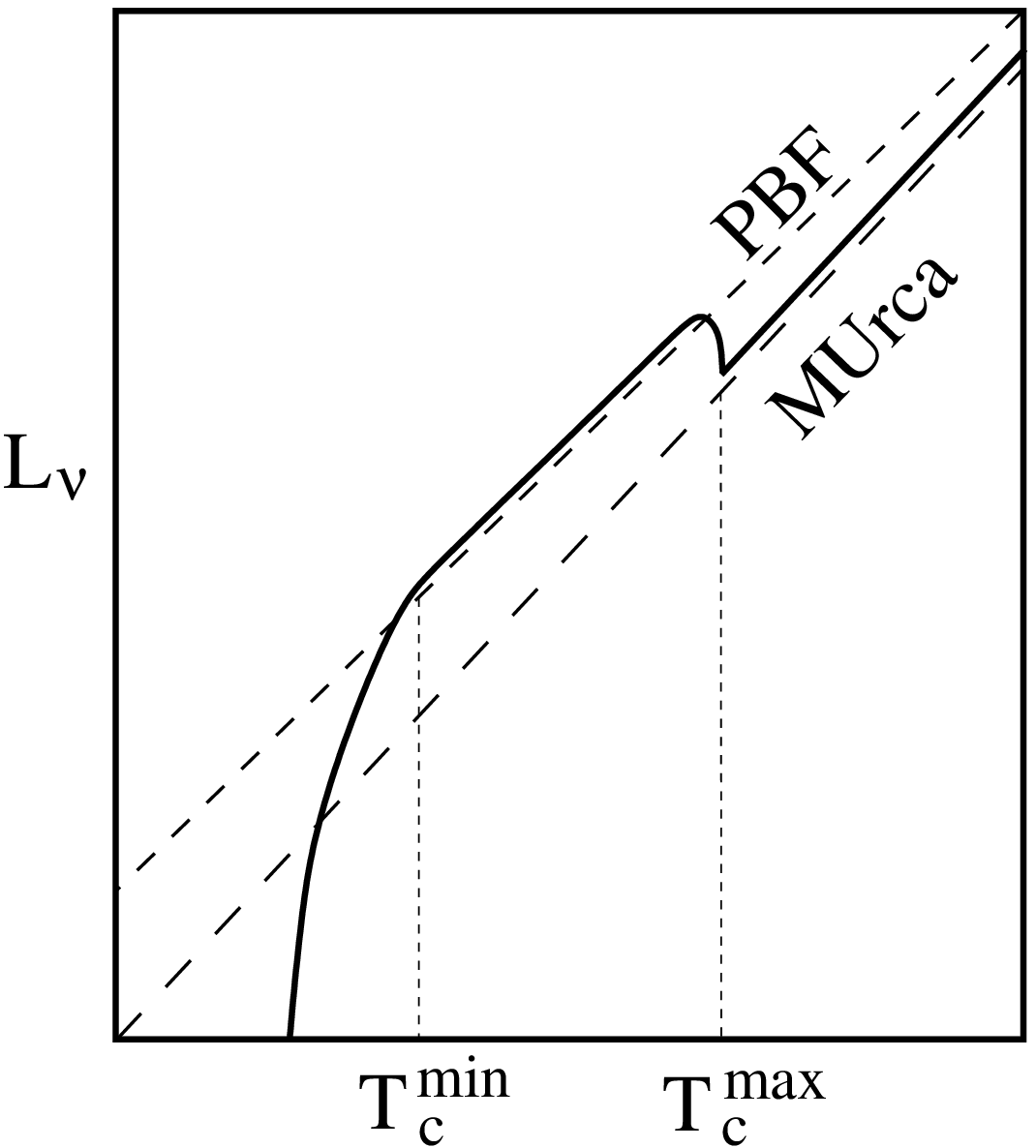}
\hspace{0.7cm}
\includegraphics[height=.20\textheight]{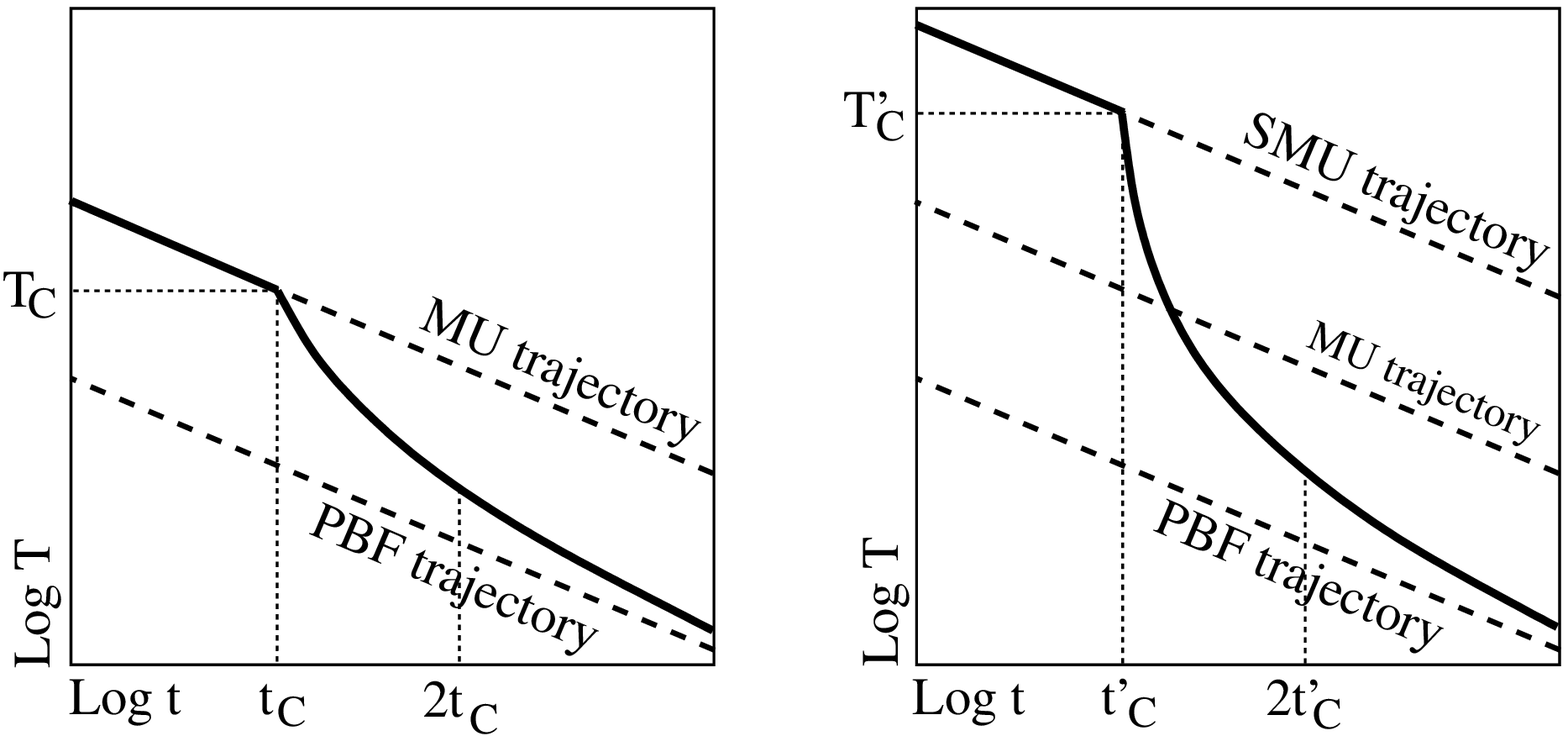}
\end{center}
\caption{See text for description. Figure from \cite{Page:2009ly} and \cite{Page:2011dz}.}
\label{Fig:Trajectories}
\end{figure*}

The constituent whose pairing has the major effect is the neutron fluid in the core. 
For illustrative purposes, the left panel of Fig.~\ref{Fig:Cooling_SF} employed a neutron $^3$P$_2$ gap 
chosen to maximize the differences between the three models:
the bell shaped $T_c(k_F)$ curve (see the examples in Fig.~\ref{Fig:3Gaps}) reaches a maximum of 
$T_C \equiv T_c^\mathrm{max} \simeq 5 \times 10^9$ K.
The right panel of Fig.~\ref{Fig:Cooling_SF} shows a series of models in which  the neutron $^3$P$_2$ gap $T_c(k_F)$
is rescaled by a constant factor, keeping the same shape.
The dotted curves, with low values of $T_C$, exhibit a transit of the neutron star from the warm trajectory with $T_C=0$
toward the cold trajectories, with $T_C=1$ or $2 \times 10^9$ K.
At early times when the star's core temperature $T$ is above $T_C$, neutrons are normal but 
when $T$ reaches $T_C$ the phase transition starts at some location in the star 
(see the left panel of Fig.~\ref{Fig:Trajectories} for a schematic of the evolution of $L_\nu$).
At that stage, neutrons in a thick shell go through the phase transition and the neutrino luminosity suddenly increases
due to the triggering of the PBF process.
The star then begins its transit toward a colder trajectory.
As $T$ decreases, this shell splits into two shells which slowly drift toward the lower and higher density regions away from the maximum of the bell-shaped $T_c(k_F)$ profile.
When $T$ is much below $T_c(k_F)$ everywhere in the core, $L_\nu$ decreases rapidly.
In the cases for which $T_C \simless 8 \times10^8$ K the onset of the phase transition occurs after the period of thermal
relaxation of the crust and is seen as a second phase of rapid decrease of $T_e$.
For larger values of $T_C$, this transitory rapid cooling of the core is hidden by the crust which is not yet thermally relaxed with the core.

The simple analytical solution of Eq.~(\ref{Eq:Simple_nu}) gives some insight into this transitory behavior.
When $T> T_C$, but $\ll T_0$, the star follows the asymptotic "MU trajectory", $T \sim t^{-1/6}$, 
and when $T$ reaches $T_C$, at time $t=t_C$, the neutrino luminosity suddenly increases.
Despite the complicated $T$ dependence of $\epsilon^\mathrm{PBF}$, Eq.~(\ref{Eq:Q_PBF}), the resulting luminosity,
once integrated over the entire core (also, aided the bell shape of the $T_c(k_F)$ curve), is well approximated
by a $T^8$ power law in the $T$ regime in which some thick shell of neutrons is going through the phase transition.
If we write $L_\nu^\mathrm{PBF} = f \cdot L_\nu^\mathrm{MU} = f N_9 T_9^8$, with $f \sim 10$, the solution of
Eq.~(\ref{Eq:Simple_nu}), replacing $t_0$ by $t_C$ and $T_0$ by $T_C$, gives
\be
T = \frac{T_C}{[1+f(t-t_C)/t_C]^{1/6}}
\;\;\;\;\;\; \mathrm{and}
\;\;\;\;\;\;
T = 10^9 \; \mathrm{K} (\tau_\mathrm{MU}/ft)^{1/6} \; (\mathrm{when} \; T\ll T_C) \; .
\label{Eq:Simple_PBF}
\ee
The central panel of Fig.~\ref{Fig:Cooling_SF} shows the two asymptotic MU and PBF trajectories and the transit trajectory.
Notice that in case the neutrino emission was already suppressed previously to the onset of neutron superfluidity,
e.g., by proton superconductivity with a larger critical temperature, the MU trajectory is replaced by a warmer suppressed MU
(SMU) trajectory and the initial part of the transit is much faster, as illustrated in the right panel of  Fig.~\ref{Fig:Cooling_SF}.

\section{Cassiopeia A and its Cooling Neutron Star}
\label{Sec:CasA}

\begin{figure*}[t]
\begin{center}
\includegraphics[height=.30\textheight]{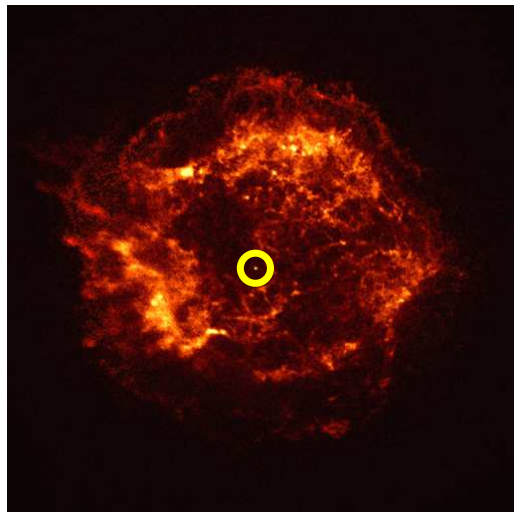}
\end{center}
\caption{The Cassiopeia supernova remnant in X-rays: first light of \texttt{Chandra}, August 1999.
(The neutron star is highlighted by the authors.) Image from \copyright NASA/CXC/SAO.}
\label{Fig:CasA}
\end{figure*}

The Cassiopeia A supernova remnant (SNR) was discovered in radio in 1947 and is the second brightest radio 
source in the sky (after the Sun).
It has since then been observed at almost all wave-lengths.
Very likely, this supernova was observed by the first Astronomer Royal, John Flamsteed 
\cite{Ashworth:1980oq} who, on August 16, 1680, when describing the stars in Cassiopeia constellation
listed the star "3 Cassiopeia" at a position almost coincident with the supernova remnant.
This star had never been reported previously, and was never seen again
until August 1999 when the first light observation of \texttt{Chandra} found a point source
in the very center of the remnant (see Fig.~\ref{Fig:CasA}).

The distance to the SNR is $3.4^{+0.4}_{-0.1}$ kpc \cite{Reed:1995kl}, and the direct observation, by the \texttt{Hubble Space Telescope}, of the remnant expansion
implies a birth in the second half of the 17$^\mathrm{th}$ century \cite{Fesen:2006tg} and 
supports  Flamsteed's observation. 
These observations  give a present age of 331 yrs for the neutron star in Cassiopeia A.
The optical spectrum of the supernova has been observed through its light echo from scattering of the
original light by a cloud of interstellar dust and  shows the supernova was of type IIb \cite{Krause:2008hc}.
The progenitor was thus a red supergiant that had lost most of its hydrogen envelope, with an estimated zero age main sequence 
(ZAMS) mass of 16 to 20 $M_\odot$
\cite{2003A&A...398.1021W,Chevalier:2003ij,van-Veelen:2009fv}
or even up to 25 $M_\odot$ in the case of a binary system \cite{Young:2006bs}.
This implies a relatively massive neutron star, i.e. likely $\simgreater 1.4 \; M_\odot$
\cite{Young:2006bs}.

The soft X-ray spectrum of the point source in the center of the SNR in Cassiopeia A is thermal, but its interpretation
has been challenging \cite{Pavlov:2004kl}.
With a known distance, a measurement of the temperature implies a measurement of the star's radius,
but spectral fits with a blackbody or a H atmosphere model resulted in an estimated radius of 0.5 and 2 km, respectively.
It was only in 2009 that a successful model was found: a non-magnetized\footnote{There is, to date, no
evidence for the presence of a magnetic field in the Cassiopeia A neutron star.}
C atmosphere, which implies a stellar radius between 8 to 18 km \cite{Ho:2009fk}.
With this model, and analyzing 5 observations of the SNR, Heinke \& Ho
\cite{Heinke:2010hc} found that the neutron star temperature had dropped by 4\% from 2000 to 2009, from
$2.12$ to $2.04 \times 10^6$ K, and the observed flux had decreased by 21\%.
The neutron star in Cassiopeia A is thus the youngest known neutron star and the first one whose cooling has been observed in real time!

\section{Superfluid Neutrons in the Core of the Cassiopeia A Neutron Star}
\label{Sec:Cool_CasA}

The \texttt{Chandra} observations of Cassiopeia A presented in the previous section, 
together with its known distance, imply that the photon luminosity of the neutron star is
\be
L_\gamma \simeq 10^{34} \; \mathrm{erg \; s}^{-1} \; .
\label{Eq:Lg_CasA}
\ee
With a measured $T_e \simeq 2 \times 10^6$ K \cite{Ho:2009fk}, 
we deduce an internal $T \simeq 4 \times 10^8$ K from Eq.~(\ref{Eq:TbTe}).
The star's total specific heat is thus $C_v \simeq 4 \times 10^{38}$ erg K$^{-1}$ (from Fig.~\ref{Fig:Cv}
or Eq.~(\ref{Eq:PL1})).
The observed $\Delta T_e/T_e \simeq$ 4\% \cite{Heinke:2010hc} 
gives for the internal temperature $\Delta T/T \simeq$ 8\% over a ten years period
since $T \sim T_e^2$.
Assuming the observed cooling corresponds to a {\em global cooling} of the whole neutron star,
its thermal energy loss is 
\be
\dot{E}_\mathrm{th} = C_V \dot{T} 
\simeq (4 \times 10^{38} \; \mathrm{erg \; K}^{-1}) \times (0.1 \; \mathrm{K \; s}^{-1} )
\simeq 4\times 10^{37}  \; \mathrm{erg \; s}^{-1}
\label{Eq:Edot_CasA}
\ee
which is 3-4 orders of magnitude larger than what it seen in $L_\gamma$ !
For a young neutron star, neutrinos are the prime candidates to induce such a large energy loss.

The cooling rate of this neutron star is so large that it must be a transitory event, which was initiated
only recently. Something critical occurred recently within this star!
"Something critical" for a cooling neutron star points toward a critical temperature, and a
phase transition is a good candidate.
The results of the previous section exhibited a phase of accelerated cooling when the neutron
$^3$P$_2$ pairing phase transition is triggered.
With a $T_C \simeq 5 \times 10^8$ K, a transitory cooling can occur at an age $\simeq 300$ yrs
as shown in the right panel of Fig.~\ref{Fig:Cooling_SF}.

\begin{figure*}[t]
\begin{center}
\includegraphics[height=.25\textheight]{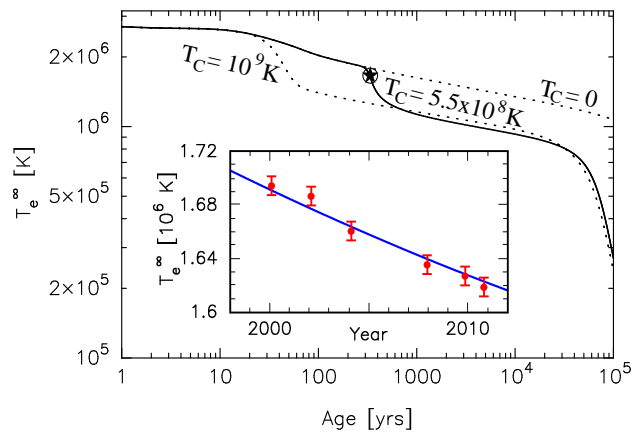}
\end{center}
\caption{
A very good fit of the rapid cooling of the neutron star in Cassiopeia A obtained assuming a recent
onset of neutron $^3$P$_2$ superfluidity and the resulting increase in neutrino emission 
from the formation and breaking of Cooper pairs.
The successful model assumes a maximum critical temperature $T_C = 5.5 \times 10^8$ K 
and the inset compares it with the six observational points, with their $1\sigma$ error bars,
from \cite{Heinke:2010hc} and \cite{Shternin:2011fu}.
The two dotted curves with no neutron superfluidity, $T_C=0$, and 
superfluidity with a higher $T_C = 1\times 10^9$ K illustrate the sensitivity to $T_C$ .
Figure adapted from \cite{Page:2011dz}.}
\label{Fig:Cool_CasA}
\end{figure*}

This interpretation of the observed rapid cooling of the neutron star in Cassiopeia  A as triggered by the
recent onset of the neutron $^3$P$_2$ superfluid phase transition and the resulting increase in
neutrino emission from the formation and breaking of pairs in the neutron superfluid was recently proposed in 
\cite{Page:2011dz} and, independently, in \cite{Shternin:2011fu}.
Models such as the ones in the right panel of Fig.~\ref{Fig:Cooling_SF} do not, however, exhibit 
a cooling rate as large as the observed one.

A second key ingredient for reproducing the observed cooling rate is illustrated in the right panel of
Fig.~\ref{Fig:Trajectories}.  The neutron star was very hot before the onset of neutron superfluidity.
This is possible in the case the protons were already in a superconducting state, which implies
that the corresponding critical temperature is significantly larger than the $T_c$ for neutrons.
That the $T_c$ for $^1$S$_0$ proton pairing is larger than the $T_c$ for neutron $^3$P$_2$ pairing
is expected from the theoretical results presented in \S~\ref{Sec:Gaps}.
However, a hot young neutron star can only be achieved in the case that the protons are superconducting 
{\em in the entire core}, so that neutrino emission form the modified Urca processes is strongly suppressed
\cite{Shternin:2011fu,Page:2011dz}.
This requirement places strong constraints on the proton $^1$S$_0$ pairing and is easier to fulfill if the 
neutron star mass is not too large as theoretical models  show that proton superconductivity
does not extend to very high densities.
A better understanding of the progenitor of Cassiopeia A, and constraining the expected neutron star mass,
is essential for this scenario to work.

A very good fit to the observations is shown in Fig.~\ref{Fig:Cool_CasA}, in which
a maximum neutron $^3$P$_2$ pairing $T_C$ of $5.5\times 10^8$ K is employed and
protons are assumed to be superconducting in the entire core with $T_c(k_F)  > 10^9$ K everywhere.
Very similar results were obtained independently in \cite{Shternin:2011fu}.

\section{Discussion and Conclusions}
\label{Sec:Conclusions}

The observation, in real time, of the cooling of a neutron star is unique and its interpretation potentially
imposes very strong constraints on the physics of ultra-dense matter.
We have here given a basic presentation of the physical principles that are involved in trying to 
understand the interior of a neutron star.
Neutrino emission processes and the likely occurrence of pairing are the two most important ingredients,
and in both cases many unsettled issues still remain.

We have presented arguments which, we hope, make it plausible that the observed 4\% temperature drop,
in a period of 10 years, of the young neutron star in the Cassiopeia A supernova remnant 
may signal the recent triggering of the phase transition of the core neutron fluid to a superfluid state.
The resulting on-going formation of $^3$P$_2$ Cooper pairs results in a strong neutrino emission
which can explain the observed rapid temperature drop of this neutron star.
An essential requirement for the cooling to be as rapid as observed is that the neutron star was
relatively hot at earlier times, and we described how this may be due to the suppression of the early
neutrino emission by proton superconductivity.
This would imply that the critical temperature, $T_c^\mathrm{p}$, for proton superconductivity is,
everywhere in the star's core, larger than $10^9$ K.
Under this condition, we found that the critical temperature for neutron superfluidity, which is density dependent,
must have a maximum value of $T_C^\mathrm{n} \simeq 5 \times 10^8$ K.
This would be the first direct evidence that neutron and proton superfluidity/superconductivity
occur at supra-nuclear densities in the core of a neutron star \cite{Page:2011dz,Shternin:2011fu}
and, further, these results would highly constrain their respective critical temperatures.

Alternatively, one could consider the observed cooling of the star in Cassiopeia A to be instead due to a 
longer thermal relaxation timescale in some layer of the star than we have assumed.
In such a case the estimate of Eq.~(\ref{Eq:Edot_CasA}), which assumes an isothermal star, does not apply.
As seen, e.g., in Fig.~\ref{Fig:Cooling_SF}, the crust thermal relaxation occurs on a time-scale of a few decades.
However, in case the crust thermal conductivity is smaller that the one employed here it may possible that
this early temperature drop be delayed till the star is 300-400 years old (see, e.g., \cite{Yakovlev:2011kl}).
Similarly, the core thermal relaxation time may be much larger than usually considered:
in case the inner core of the star cools rapidly it may take a few hundreds years for the star to become isothermal
and this can induce a rapid decrease of $T_e$,
as proposed in \cite{Blaschke:2011qa} and presented in these proceedings by D. Blaschke.
This scenario requires that the core thermal conductivity be lower than usually considered, by a factor
of a few, and also implies that neutrons do not form a superfluid until the star is much colder.

The neutron superfluid scenario we have presented fits within the {\em minimal cooling} scenario 
\cite{Page:2004zr} and is compatible with observations of other cooling neutron stars \cite{Page:2009ly}.
The very similar model of  \cite{Shternin:2011fu} also successfully passes the same test as well as the
opposite core relaxation scenario of \cite{Blaschke:2011qa}.
More work is required to confront these possibilities with other facets of the neutron star phenomenology.

\acknowledgments
DP's acknowledges support by grants from UNAM-DGAPA, \# PAPIIT-113211, and Conacyt, CB-2009-01, \#132400,
and thanks the organizers of the {\em XXXIV Brazilian Workshop on Nuclear Physics} for the generous invitation. 
MP and JML acknowledge research support from  
the U.S. DOE grants DE-FG02-93ER-40756 and DE-AC02-87ER40317, respectively.
AWS is supported by Chandra grant TM1-12003X, by NSF PHY grant 08-22648, 
by NASA ATFP grant NNX08AG76G, and by DOE grant DE-FG02-00ER41132.



\providecommand{\bysame}{\leavevmode\hbox to3em{\hrulefill}\thinspace}
\providecommand{\MR}{\relax\ifhmode\unskip\space\fi MR }
\providecommand{\MRhref}[2]{%
  \href{http://www.ams.org/mathscinet-getitem?mr=#1}{#2}
}
\providecommand{\href}[2]{#2}


\end{document}